\documentclass[pre,superscriptaddress,showpacs,amsmath,amssymb]{revtex4}
\usepackage{graphicx}
\usepackage{graphicx,amsfonts}
\usepackage{epsfig,amsmath}
\usepackage{verbatim}
\begin{document}

\newcommand{\ArcTan}
{\operatorname{ArcTan}}
\newcommand{\ArcSin}
{\operatorname{ArcSin}}

\newcommand{\beq}{\begin{equation}}
\newcommand{\eeq}{\end{equation}}
\newcommand{\bea}{\begin{eqnarray}}
\newcommand{\eea}{\end{eqnarray}}

\title{Real Roots of Random Polynomials and Zero Crossing
  Properties of Diffusion Equation.}

\author{Gr{\'e}gory Schehr}
%\email{schehr@lpt.ens.fr}
\affiliation{Laboratoire de Physique Th\'eorique (UMR du
  CNRS 8627), Universit\'e de Paris-Sud, 91405 Orsay Cedex,
  France}
\author{Satya N. Majumdar}
\affiliation{Laboratoire de Physique Th\'eorique et et Mod\`eles
  Statistiques, Universit\'e Paris-Sud, B\^at. 100, 91405 Orsay Cedex,
France}

\date{\today}

\begin{abstract}
We study various statistical properties of real roots of three
different classes of random 
polynomials which recently attracted a vivid interest in the context of probability theory
and quantum chaos. We first focus on gap probabilities on the real axis, {\it i.e.} the probability
that these polynomials have no real root in a given interval. For generalized Kac polynomials, indexed
by an integer $d$, of large degree $n$, one finds that the probability of no real root in the interval $[0,1]$ decays 
as a power law $n^{-\theta(d)}$ where $\theta(d) > 0$ is the persistence exponent of the diffusion equation with random
initial conditions in spatial dimension $d$. For $n \gg 1$ even,  
the probability that they have no real root on the full real axis decays like $n^{-2(\theta(2)+\theta(d))}$. For Weyl polynomials
and Binomial polynomials, this probability decays respectively like $\exp{(-2\theta_{\infty}} \sqrt{n})$ and $\exp{(-\pi \theta_{\infty} \sqrt{n})}$
where $\theta_{\infty}$ is such that $\theta(d) = 2^{-3/2} \theta_{\infty} \sqrt{d}$ in large dimension $d$. We also show that the probability that such polynomials
have exactly $k$ roots on a given interval $[a,b]$ has a scaling form given by $\exp{(-N_{ab} \, \tilde \varphi(k/N_{ab}))}$ where 
$N_{ab}$ is the mean number of real roots in $[a,b]$ and $\tilde \varphi(x)$ a universal scaling function. 
We develop a simple Mean Field (MF) theory reproducing qualitatively these scaling behaviors, and improve systematically this MF approach using the method of persistence with partial survival, which in some cases yields exact results.  Finally, we show that the probability density function of the largest
absolute value of the real roots has a universal algebraic tail with exponent ${-2}$.  These analytical results are confirmed by detailed 
numerical computations. Some of these results were announced in a recent letter [G.~Schehr and S.~N.~Majumdar, Phys. Rev. Lett. {\bf 99}, 060603 (2007)].
\end{abstract}

\maketitle

\section{Introduction}

 Despite several decades of research, understanding the zero crossing
 properties of non-Markov stochastic processes remains a 
challenging issue. Among them, the persistence probability $p_0(t)$ received a particular
attention, especially in the context of many-body non-equilibrium
 statistical   
physics, both analytically \cite{satya_review} as well as
experimentally \cite{marcos, persist_exp_froth, persist_exp_steps,
  persist_diffusion_exp}.  
The persistence $p_0(t)$ for a time dependent stochastic process with
 zero mean is defined as the probability that it has not changed sign up to time $t$. In various 
physical situations, $p_0(t)$ has a power law tail $p_0(t) \sim
t^{-\theta}$ where $\theta$ turns out to be a non-trivial
exponent whenever the stochastic process under study has 
a non Markovian dynamics. One such example is the diffusion, or heat
 equation in space dimension $d$ where a scalar field $\phi({\mathbf
 x},t)$ evolves 
according to the deterministic equation 
\begin{eqnarray}
\partial_t \phi({\mathbf x},t) = \nabla^2 \phi({\mathbf x},t) \, ,
\label{diff_eq} 
\end{eqnarray}
with {\it random} initial conditions $\phi({\mathbf x},t=0) = \psi({\mathbf x})$
where $\psi({\mathbf x})$ is a Gaussian random field of zero mean 
with delta correlations $[\psi({\mathbf x}) \psi({\mathbf
  x'}) ]_{\rm ini} = \delta^{d}({\mathbf x}-{\mathbf x'})$. We use the
notation $[...]_{\rm ini}$ to denote 
an average over the initial condition.  For a system of linear size $L$, the persistence $p_0(t,L)$
is the probability that $\phi({\mathbf x},t)$, at some fixed point ${\mathbf x}$ in space, does not change
sign up to time $t$.  The initial condition being (statistically)
invariant under translation in space, this probability does not depend 
on the position ${\mathbf x}$. In the scaling limit
$t \gg 1$, $L \gg 1$ keeping the ratio $t/L^2$ fixed, it was 
found in Ref.~\cite{persist_diffusion} that $p_0(t,L)$ takes the scaling form
\begin{equation}
p_0(t,L) \propto L^{-2\theta(d)} h (L^2/t) \;,
\label{fss1}
\end{equation}
where $h(u) \sim c^{\rm st}$, a constant independent of $L$ and $t$,
for $u \ll  
1$ and $h(u) \propto u^{\theta(d)}$ for $u \gg 1$ where $\theta(d)$ is
a $d$-dependent exponent. This implies that in the $L \to \infty$ limit,
$p_0(t) \equiv p_0(t,L\to \infty) \sim t^{-\theta(d)}$ for large
$t$. It was shown in
Ref.~\cite{persist_diffusion} that the probability ${\cal P}_0(T)$
that a Gaussian stationary process (GSP) 
with zero mean and correlations $[\cosh(T/2)]^{-d/2}$ decays for large
$T$ as ${\cal P}_0(T) \sim \exp{[-\theta(d) T]}$ where $\theta(d)$ is
the same as the persistence exponent in diffusion equation. 
This exponent $\theta(d)$ was measured 
in numerical simulations \cite{persist_diffusion, newman_diffusion},
yielding for instance $\theta_{\rm sim}(1) = 0.12050(5)$, 
$\theta_{\rm sim}(2) = 0.1875(1)$. The case of dimension $d=1$ is
particularly interesting because $\theta(1)$ was determined 
experimentally using NMR techniques to measure the magnetization of
spin polarized Xe gas \cite{persist_diffusion_exp}, 
 yielding $\theta_{\rm exp}(1) = 0.12$ in good agreement with
 numerical simulations. In the limit of large dimension $d$, which will be 
of interest in the following, one can show that $\theta(d) = 2^{-3/2}
\theta_{\infty} \sqrt{d}$ where 
$\theta_{\infty}$ is the decay constant associated with the no zero
crossing probability of the  GSP with Gaussian correlations
$\exp{(-T^2/2)}$, which was studied in the past by engineers, in 
particular in the context of fading of long-wave radio signals (see for
instance Ref.~\cite{palmer}).

A seemingly unrelated topic concerns the study of random algebraic
equations which, since the first work by Bloch and P\'olya \cite{bloch} in the 30's, 
has now a long story \cite{bharucha, farahmand}. Recently
it has attracted a renewed interest  
in the context of probability and number theory \cite{edelman} as well
as in the field of quantum chaos \cite{bogomolny}. In a recent letter
\cite{us_prl}, we have established a close 
connection between zero crossing properties of the diffusion equation
with random initial  
conditions~(\ref{diff_eq}) and the real roots of real
random polynomials ({\it i.e.} polynomials with real random coefficients). In
Ref.~\cite{us_prl}, 
we focused on a class of real random polynomials $K_n(x)$ of degree
$n$, the so called generalized Kac polynomials, indexed by an integer
$d$ 
\begin{eqnarray} 
K_n(x) = a_0 + \sum_{i=1}^{n} a_i \; i^{\tfrac{d-2}{4}} x^i \;.
\label{def_kac_poly}  
\end{eqnarray} 
Here, and in the following, $a_i$'s are independent real Gaussian random variables of zero
mean and unit variance $\langle a_i a_j \rangle = \delta_{ij}$ where
we use the notation $\langle ... \rangle$ to denote an average
over the random coefficients $a_i$. In the
case of $d=2$, these polynomials reduce to the standard Kac polynomials
\cite{kac_1}, which have been extensively studied in the past (see for
instance Ref.~\cite{edelman} for a recent review). In that case, we
will see below that the 
statistics of real roots of $K_n(x)$ is identical in the $4$
sub-intervals $[-\infty, -1], [-1,0], [0,1]$ and
$[1,+\infty]$. Instead, for $d \neq 2$, which was studied in
Ref.~\cite{das}, the statistical behavior 
of real roots of $K_n(x)$ depend on $d$ in the inner intervals,
while it is identical to the case $d=2$ in the outer
ones. Focusing on the interval $[0,1]$, we asked the question : what is 
the probability $P_0([0,x],n)$, $0 < x < 1$, that $K_n(x)$ has no
real root in the interval $[0,x]$ ? Such probabilities were often
studied in the context of random matrices, where they are known as
{\it gap probabilities} \cite{mehta} and in a recent work \cite{dembo},
Dembo {\it et al.} showed that, for random polynomials $K_n(x)$ with
$d=2$, $P_0([0,1],n) \propto n^{-\zeta(2)}$ where the exponent
$\zeta(2) = 0.190(8)$ was computed numerically. In Ref.~\cite{us_prl},
by mapping these two
random processes (\ref{diff_eq}) and (\ref{def_kac_poly}) 
onto the same GSP, we showed that  
in the limit $1-x \ll 1 \ll n$ keeping $n(1-x)$ fixed, one has
(similarly to Eq. (\ref{fss1})) 
\beq
P_0([0,x],n) \propto n^{-\theta(d)} h^-(n(1-x)) \;,\label{persist_poly}
\eeq
with $h^-(y) \to 1$ for $y \ll 1$ and $h^-(y) \sim y^{\theta(d)}$ for
$y \gg 1$, yielding in particular $P_0([0,1],n) \propto n^{-\zeta(d)}$
thus identifying 
$\zeta(d) = \theta(d)$. We then extended our study to the
probability $P_k([0,1],n)$ that generalized Kac polynomials  have exactly
$k$ real roots in $[0,1]$ and we showed that it has an unusual scaling form
(for large $k$, large $n$, but keeping the ratio $k/\log n$ fixed)
\begin{eqnarray}
P_k([0,1],n) \propto n^{-\tilde\varphi\left(\tfrac{k}{\log n} \right)} \;,
\label{rpscaling1}
\end{eqnarray}           
where $\tilde \varphi(y)$ is a large deviation function, with $\tilde
\varphi(0) = 
\theta(d)$. In both cases, our numerical analysis suggested that
$h^-(y)$ and $\tilde \varphi(y)$ are
universal in the sense that they are independent of the 
distribution of $a_i$ provided $\langle a_i^2 \rangle$ is finite. The
purpose of the present paper is twofold : (i) we will give detailed
derivations of the results announced in Ref.~\cite{us_prl} together
with some new results, like the distribution of the largest real root, for
generalized Kac polynomials  $K_n(x)$~(\ref{def_kac_poly}), (ii) we
extend these results (\ref{persist_poly}, \ref{rpscaling1}) to two
other classes of random polynomials which were recently considered
in the literature. First we will study Weyl polynomials $W_n(x)$
defined as   
\begin{eqnarray}
W_n(x) = \sum_{i=0}^{n} a_i \; \frac{x^i}{\sqrt{i !}} \;. \label{def_weyl_poly}
\end{eqnarray}
Recently, the distribution of complex zeros of Weyl polynomials with
complex coefficients were 
observed experimentally in a degenerate rotating quasi-ideal atomic
Bose gas \cite{castin_complex_exp}. Here we will focus instead on the
real roots of such polynomials (\ref{def_weyl_poly}) with real coefficients. Besides, we will consider binomial
polynomials $B_n(x)$ defined as 
\begin{eqnarray}
B_n(x) = \sum_{i=0}^{n} a_i \; \sqrt{ n \choose i } x^i
\;. \label{def_binom_poly}   
\end{eqnarray}
As is pointed out by Edelman and Kostlan \cite{edelman}, "this
particular random polynomial is probably the 
more natural definition of a random polynomial". In the literature,
they are sometimes called $SO(2)$ random polynomials because their
$m$-point joint probability distribution of zeros is $SO(2)$ invariant
for all $m$ \cite{bleher}. We will show below that the gap
probabilities for these classes of random polynomials (\ref{def_weyl_poly}, \ref{def_binom_poly}) are closely
related to the persistence probability for 
the diffusion equation in the limit of large dimension. Our main results, together with the layout of the paper, are summarized below.

\begin{enumerate}
\item{In section II, we briefly recall
the main properties of the persistence probability, $p_0(t,L)$, for the
diffusion equation with random initial conditions. In section II-A, we recall the finite size scaling for $p_0(t,L)$ in dimension $d$ whereas
in section II-B, we focus on the limit $d \to \infty$. }
\item{Section III is devoted to real random polynomials, where our main results are presented. In section III-A, we
present a detailed study of the density of real roots for these three
classes of polynomials, which turns out to behave quite differently in all the the three cases
under investigations (\ref{def_kac_poly},~\ref{def_weyl_poly},~\ref{def_binom_poly})~. In section III-B, we will turn to 
the analysis of gap probabilities, which we will first analyse from the point of view of two-point correlations. Next, we 
will present a mean field approach, or Poissonian approximation, which neglects the correlations between the real roots of these polynomials, 
to compute the gap probabilities. We will further show how this mean field approximation can be systematically 
improved using the persistence probability with partial survival~\cite{satya_partial}, which in some cases even yields exact results. In particular we show that
the probability $q_0(n)$ that these polynomials have no root on the full real axis is given by 
\bea
&&q_0(n) \sim n^{-2(\theta(d)+\theta(2))} \hspace*{1.55cm}  {\rm for \;} K_n(x) \;,\nonumber \\
&&q_0(n) \sim \exp{(-2\theta_{\infty}} \sqrt{n}) \hspace*{1.3cm}  {\rm for \;} W_n(x) \;, \nonumber \\
&&q_0(n) \sim \exp{(-\pi \theta_{\infty} \sqrt{n})} \hspace*{1.26cm} {\rm for \;} B_n(x) \;.
\eea
In section III-C, we will then generalize this study to the probability that these polynomials have exactly $k$ real roots on a given real interval.
Extending the results obtained in Ref.~\cite{us_prl} for $K_n(x)$ like in Eq. (\ref{rpscaling1}), to Weyl
and Binomial polynomials, we will
show that the probability $q_k(n)$ that  $W_n(x)$ and $B_n(x)$ have exactly $k$ roots on the full real axis has 
a scaling form 
(for large $k$, large $n$, but keeping the ratio $k/\sqrt{n}$ fixed) given by
\bea\label{rpscaling2intro}
q_k(n) \sim \exp{[-\sqrt{n} \tilde \varphi\left({k}/{\sqrt{n}} \right)]} \;,
\eea
where $\tilde \varphi(y)$ is a large deviation function, which depends on the polynomials under consideration 
$W_n(x)$ or $B_n(x)$. We will also show that these scaling forms  in Eq. (\ref{rpscaling1}, \ref{rpscaling2intro}) can be
qualitatively described by the aforementioned mean field approximation. To end up, we study in section III-D the probability
density (p.d.f.) $p_{\rm max}(x)$ of the largest absolute value of the real roots and obtain the exact asymptotic result
\begin{eqnarray}
p_{\rm max}(x) \propto \frac{1}{x^2} \;, \; x \gg 1 \;,
\end{eqnarray}
for all the three classes of random polynomials under investigation. All our analytical results will be verified by numerical computations and some details of 
the analytical computations involved in this section have been left in Appendices A,B, C, D and E.
}
\item{Finally section IV contains our conclusions and perspectives.}
\end{enumerate}

\section{A brief overview on persistence for diffusion equation in dimension
  $d$}  

\subsection{Persistence exponent $\theta(d)$ and finite size scaling}

We consider a scalar field $\phi({\mathbf x},t)$ in a $d$-dimensional space
which evolves in time under the diffusion equation (\ref{diff_eq}).
For a system of linear size $L$, the solution of 
the diffusion equation in the bulk of the system is 
\begin{eqnarray}
\phi({\mathbf x},t) = \int_{|{\mathbf y}| \leq L} d {\mathbf y} \; {\cal
  G}({\mathbf x}-{\mathbf y},t) \; \psi({\mathbf y}) \quad, \quad  {\cal
  G}({\mathbf x}) = (4 \pi t)^{-d/2} \exp{(-{\mathbf x}^2/4t)} \, ,
\label{sol_eqdiff} 
\end{eqnarray} 
where $\psi(\mathbf x) = \phi(\mathbf x,0)$ is the initial
uncorrelated Gaussian field. Since Eq. (\ref{sol_eqdiff}) is linear,
$\phi({\mathbf x},t)$ is a Gaussian variable for all time $t \geq 
0$. Therefore its zero crossing properties are completely determined
by the two time correlator $[ \phi({\mathbf x},t) \phi({\mathbf
  x},t')]_{\rm ini}$.  
To study the persistence probability $p_0(t,L)$ it is customary to
study the normalized 
process $X(t) = \phi({\mathbf x},t)/[ \phi({\mathbf x},t)^2]_{\rm ini}^{1/2}$
\cite{satya_review}. Its autocorrelation 
function $a(t,t') = [ X(t) X(t') ]_{\rm ini}$ is computed
straightforwardly from the solution in Eq. (\ref{sol_eqdiff}). One obtains
$a(t,t' )\equiv a(\tilde t, \tilde t')$ with $\tilde t= t/L^2$,
$\tilde t' = t'/L^2$ and 
\begin{eqnarray}\label{attprime}
a(\tilde t,\tilde t') = 
\begin{cases}
\left(\frac{4 \tilde t \tilde t'}{(\tilde t+\tilde t')^2}\right)^{d/4} \quad, \quad \tilde t,\tilde t' \ll 1 \\
1 \quad, \quad \tilde t,\tilde t'\gg 1 \, .
\end{cases}
\end{eqnarray}
We first focus on the time regime $\tilde t, \tilde t' \ll 1$.  In terms of  
logarithmic time variable $T = \log \tilde t$, $X(T)$ is a GSP with correlator 
\begin{eqnarray}
a(T,T') \equiv a(T-T') = [{\rm cosh}(|T-T'|/2)]^{-d/2} \; ,
\label{correl_gsp_diff} 
\end{eqnarray}
 which decays exponentially for large $|T-T'|$. Thus the persistence probability $p_0(t,L)$,
for $t\ll L^2$, reduces to the computation of the probability ${\cal P}_0(T)$
of no zero 
crossing of $X(T)$ in the interval $[0,T]$. It is well known \cite{slepian}
that if $a(T) < 1/T$ at large $T$ then ${\cal P}_0(T)\sim \exp[-\theta T]$
for large $T$ where the decay constant $\theta$ depends
on the full stationary correlator $a(T)$. Reverting back to 
the original time $\tilde t=e^T$, one
finds $p_0(t,L) \sim 
t^{-\theta(d)}$, for $t \ll L^{2}$. In the opposite limit $t \gg L^2$, one has
$p_0(t,L) \to A_L$, a constant which depends on $L$. These two
limiting behaviors 
of $p_0(t,L)$ can be combined into a single finite size scaling form
as in Eq. (\ref{fss1}). 

Despite many efforts, there exists no exact result for $\theta(d)$. However various approximation methods have been developed to estimate it.
One of the most powerful is the so called Independent Interval Approximation (IIA)~\cite{IIA}, which assumes the statistical
independence of the intervals between successive zeros of $\phi(\mathbf{x},t)$. This gives {\it e.g.} $\theta_{\rm IIA}(1)~=~0.1203...$, 
$\theta_{\rm IIA}(2)~=~0.1862...$ \cite{persist_diffusion} in
remarkable agreement with numerical simulations. A more systematic
approach is via persistence with partial survival
\cite{satya_partial}, which we will use below (see section III-B). An
alternative systematic  
approach consists in performing a small $d$ expansion \cite{hilhorst}
yielding $\theta(d) = d/4 - 0.12065...d^{3/2}+...$, which would
certainly require higher order terms to make it numerically
competitive. Yet another systematic approach is a series expansion
introduced in the context of "discrete time persistence", yielding
results for $\theta(d)$ which are in very good agreement with
numerical simulations~\cite{ehrhardt}.

\subsection{Persistence in the limit of large dimension $d$}

As we will see later, some statistical properties of the real roots of
the polynomials $W_n(x)$ (\ref{def_weyl_poly}) and $B_n(x)$ (\ref{def_binom_poly}) turn out to be related to the
statistics of zero crossings of the diffusion equation in the limit of
large dimension $d$. To study the persistence probability in that limit one 
performs a rescaling of the $T$ variable in
Eq. (\ref{correl_gsp_diff}), $T = 2^{3/2} \tilde T/\sqrt{d}$
such that  
\begin{eqnarray}
a(T-T') = a \left( 2^{3/2}\frac{\tilde T- \tilde T'}{\sqrt{d}}   \right)
\sim \exp{\left[-\frac{1}{2}(\tilde T- \tilde T')^2\right]} \;, \; d \gg 1 \,.\label{correlator_larged} 
\end{eqnarray}
Therefore in the limit of large dimension $d$, one has 
$\theta(d) = 2^{-3/2} \theta_{\infty} \sqrt{d}$ where
$\theta_{\infty}$ is the decay constant associated with the no zero
crossing probability of the  GSP with correlator
$\exp{[-\frac{1}{2}(T-T')^2]}$. Even in that limit, there is no
exact result for $\theta_\infty$. However, it can be approximately
estimated using IIA \cite{IIA}, yielding
$\theta_{\infty, {\rm IIA}} = 0.411497...$ \cite{persist_diffusion} in very
good agreement with numerical simulations $\theta_{\infty, {\rm sim}}
= 0.417(3)$ \cite{newman_diffusion}.   

\section{Random polynomials}

We now focus on statistical properties of the real roots of
random polynomials, extending our previous study presented in Ref.~\cite{us_prl}. 
%For a given random polynomial ${Q}_n(x) = \sum_{i=0}^n b_i x^i$ we
%define the polynomial $\overline{Q}_n(x)$ as
%\begin{eqnarray}
%\overline{Q}_n(x) = \sum_{i=0}^n b_{n-i} x^i \;,\label{define_bar}
%\end{eqnarray}
%and in the following we will consider $\overline{K}_n(x),
%\overline{W}_n(x)$ and $\overline{B}_n(x)$.
Being Gaussian processes, the statistical properties of these polynomials are
determined by the $2$-point correlators $C_n (x,y)$, given by 
\begin{eqnarray}
C_n (x,y) &=& \langle K_n(x) K_n(y) \rangle = 1 + \sum_{i=1}^{n}
i^{\tfrac{d-2}{2}} \; (xy)^i  \; \hspace*{3cm} {\rm for \; Kac \; polynomials} \;,   \label{correl_kac} \\
C_n (x,y) &=& \langle W_n(x) W_n(y) \rangle = \sum_{i=0}^n \frac{(xy)^i}{i
  !}  \; \hspace*{4.25cm}  {\rm for \; Weyl \; polynomials} \;, \label{correl_weyl} \\
C_n(x,y) &=& \langle B_n(x) B_n(y) \rangle = \sum_{i=0}^n {n
  \choose i} (x y)^i  = (1+xy)^n \; \hspace*{1.8cm}{\rm for \; Binomial \; polynomials} \;. \label{correl_binom}
\end{eqnarray}
For simplicity, we chose the same notation $C_n(x,y)$ for the three classes of polynomials under study, and we will do so for other quantities. In the following, these three polynomials will be treated 
separately so this should not induce any confusion. For later purposes it is 
convenient to introduce the normalized correlator $\hat C_n (x,y)$ with 
\begin{eqnarray}
\hat C_n(x,y) = \frac{C_n(x,y)}{\sqrt{C_n(x,x)
    C_n(y,y)}} \;.
\end{eqnarray}
Notice that $\hat C_n(x,y) = \hat C_n(1/x,1/y)$ for Kac polynomials $K_n(x)$ with $d=2$ and for Binomial polynomials $B_n(x)$.

\subsection{Density and mean number of real roots}

Let us denote $\lambda_1, \lambda_2, ..., \lambda_p$ the $p$ real
roots (if any) of one of these random polynomials in
Eq. (\ref{def_kac_poly},~\ref{def_weyl_poly},~\ref{def_binom_poly}). The 
mean density of real roots $\rho_n(x)$ is given by
\beq
\rho_n(x) = \sum_{i=1}^p \langle \delta(x-\lambda_i) \rangle = 
\langle |K_n'(x)|\delta(K_n(x)) \rangle \;, \hspace*{1cm} {\rm for \; Kac \; polynomials} \;,\label{def_density}
\eeq
and similarly for Weyl polynomials $W_n(x)$ and Binomial polynomials $B_n(x)$. Under this form (\ref{def_density}), one observes that the computation of 
the mean density involves the joint distribution of the polynomial
$K_n(x)$ and its derivative $K'_n(x)$ which is simply a bivariate
Gaussian distribution. Thus computing $\rho_n(x)$ involves a double
integration of a bivariate Gaussian distribution. This can be easily
performed to obtain the following well known result
\begin{eqnarray}\label{def_density_inter}
\rho_n(x) = \frac{\sqrt{c_n(x) (c_n'(x)/x + c_n''(x)) - [c_n'(x)]^2   }}{2 \pi c_n(x)} \;, \; c_n(x) = C_n(x,x) \;.
\end{eqnarray}
This formula (\ref{def_density_inter}) can be written in a very compact way \cite{edelman} :
\begin{eqnarray}
\rho_n(x) = \frac{1}{\pi} \sqrt{\partial_u \partial_v \log C_n(u,v)}
\bigg 
|_{u=v=x} \;.\label{ek_formula}
\end{eqnarray}   
For these different polynomials in
 Eq. (\ref{def_kac_poly},~\ref{def_weyl_poly},~\ref{def_binom_poly}), we  
 will be interested in the number of real roots on a given interval $[a,b]$, which we will
denote $N_n[a,b]$. Being a random variable, we will focus on its moments 
$\langle N^k_n[a,b] \rangle$, with $k \in \mathbb{N}$. In particular, one has from the definition of $\rho_n(x)$
in Eq. (\ref{def_density})
\beq
\langle N_n[a,b] \rangle = \int_a^b \rho_n(x) \, dx \;,
\eeq
and higher cumulants will be considered below.

\subsubsection{Generalized Kac polynomials}

One remarkable property of the generalized Kac polynomials  $K_n(x)$ is that,
in the large $n$ limit, the roots in the complex plane tend to
accumulate close to the unit circle centered at the origin. 
\begin{figure}[h]
\begin{minipage}{0.5\linewidth}
\includegraphics[angle=-90,width=\linewidth]{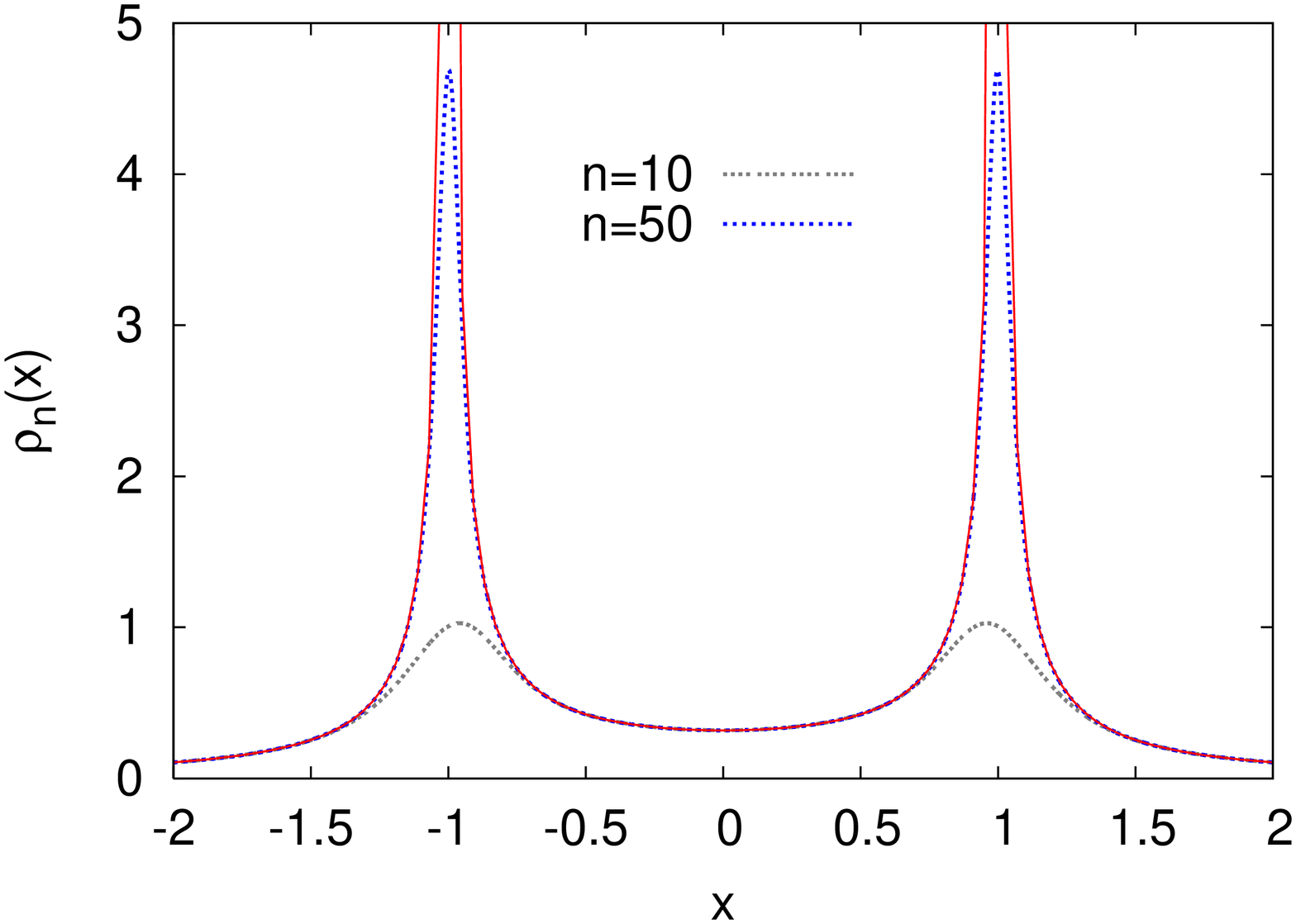}
\end{minipage}\hfill
\begin{minipage}{0.5\linewidth}
\includegraphics[angle=-90,width=\linewidth]{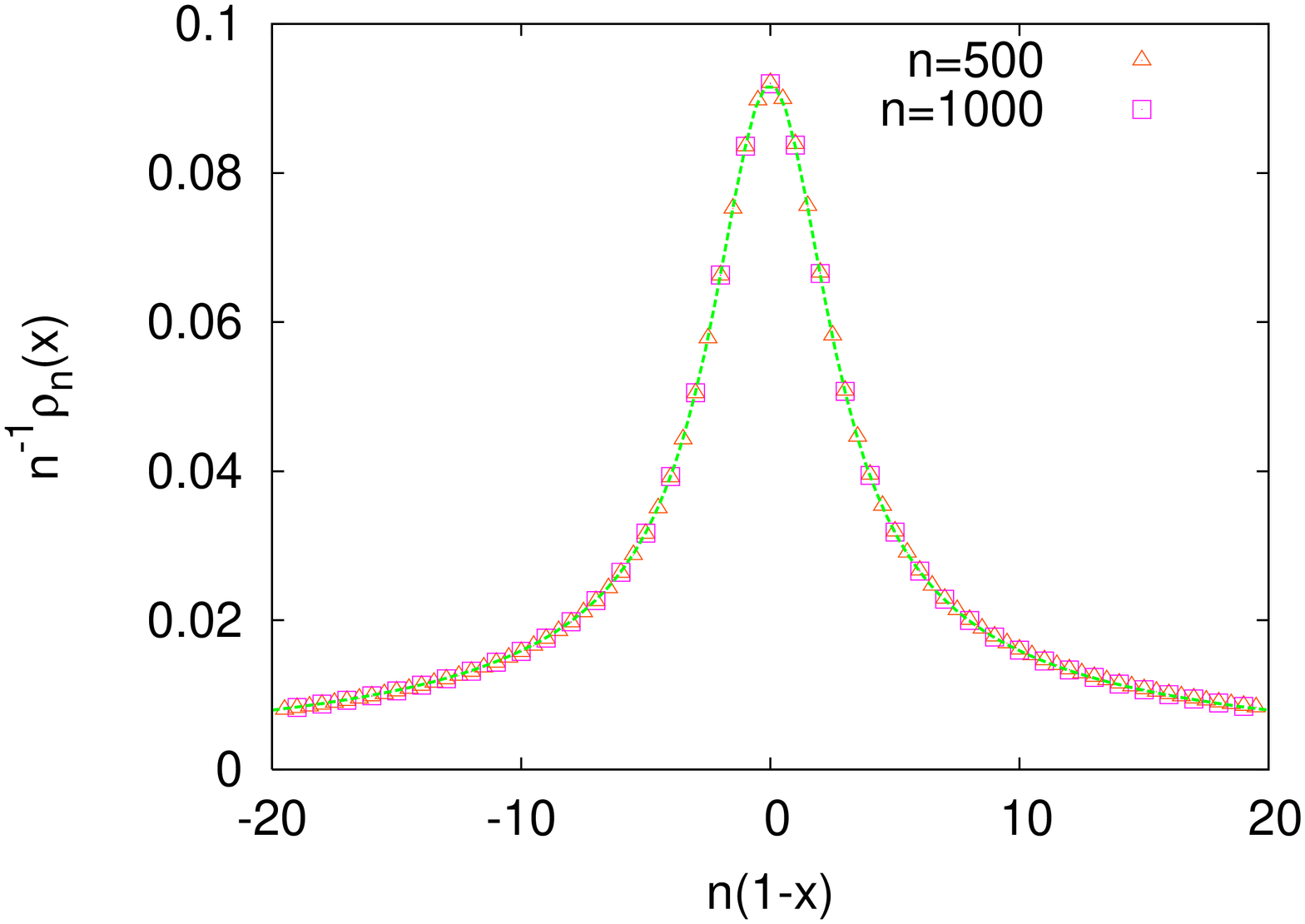}
\end{minipage}
\caption{{\bf Left} : Mean density of real roots $\rho_n(x)$ for
  Kac polynomials $K_n(x)$ (\ref{def_kac_poly}) and $d=2$ as a
  function of $x$ for different values of $n=10, 50$ (dotted
  lines). The solid line is the analytic expression $\rho_\infty(x)$
  in Eq.~(\ref{examples_density}) for $|x| < 1$ and in
  Eq.~(\ref{density_kac_outer}) for $|x|>1$. {\bf
  Right} : Plot of $n^{-1} \rho_n(x)$ as a function of $n(1-x)$ for
  $n=500, 1000$ (and thus 
  $x$ close to $\pm 1$). The dotted line is the function $\rho^K(y)$ in Eq.~(\ref{scaled_density_kac_2d}). There is no fitting
  parameter.}\label{Fig1} 
\end{figure}
 In the left panel of Fig. \ref{Fig1}, we show a plot of the density
 of real roots $\rho_n(x)$ for $d=2$ computed  
from Eq. (\ref{ek_formula}) for different values of $n=10$ and
 $50$. In the large $n$ limit,  
one clearly sees that the real roots of such polynomials are
 concentrated around $x = \pm 1$, where the density is diverging.  
This can be seen by computing $\rho_n(\pm 1)$ from Eq. (\ref{ek_formula})
\begin{eqnarray}
\rho_n(\pm 1) = \frac{1}{\pi}
\frac{\left(  (1+{\cal H}(n,1-d/2)) {\cal
    H}(n,-d/2-1)  - {\cal H}(n,-d/2)  ^2      \right)^{\tfrac{1}{2}}}{
  1 + {\cal H}(n,1-d/2)} \propto \frac{2 n}{\pi (d+2)} \sqrt{\frac{d}{d+4}} \;,
\label{divergence_at_unity} 
\end{eqnarray}
where ${\cal H}(n,r) = \sum_{k=1}^n k^{-r}$ is a generalized harmonic
number \cite{grad}. To obtain the asymptotic behavior in the large $n$ limit of 
the above equation (\ref{divergence_at_unity}) we used  ${\cal
  H}(n,r) \propto n^{1-r}/(1-r)$, for large $n$ and $r<1$.

Away from these singularities, $\rho_n(x)$ has a good limit when $n \to
\infty$. However, one has to treat separately the cases $|x| <
1$ and $|x| > 1$. For $|x| < 1$, the
calculation is straightforward because $C_{n}(x,y)$ in
Eq. (\ref{correl_kac}) has a good limit $n \to \infty$ when $x,y <
1$. This yields 
\begin{equation}
\rho_{\infty}(x) = \frac{\left[{
    {\rm Li}_{-1-d/2}(x^2)(1+{\rm Li}_{1-d/2}(x^2)) - 
    {\rm Li}^2_{-d/2}(x^2)  }\right]^{\tfrac{1}{2}}} 
{\pi|x|(1+{\rm Li}_{1-d/2}(x^2))} \quad, \quad |x| < 1 \;, \label{mean_density}
\end{equation} 
where ${\rm Li}_n(z) = \sum_{i=1}^\infty z^i/i^n$ is the polylogarithm
function \cite{grad}. In particular, one has $\rho_{\infty}(0) = 1/\pi$ for all 
$d$, and $\rho_{\infty}(x) \sim (d/2)^{\tfrac{1}{2}}(2\pi((1-x)))^{-1}$
for $x \to 1^-$. For instance, one has for $|x|<1$
\begin{eqnarray}\label{examples_density}
\rho_{\infty}(x) = \frac{1}{\pi (1-x^2)} \quad {\rm in} \quad d=2
\quad, \quad \rho_{\infty}(x) = \frac{1}{\pi (1-x^2)}
\sqrt{\frac{x^8+2x^6-4x^4+2x^2+1}{x^8-2x^6+3x^4-2x^2+1}} \quad {\rm in}
\quad d=4 \;.
\end{eqnarray}
For $|x|>1$, the analysis is different because the
correlator $C_n(x,y)$ in Eq. (\ref{correl_kac}) does not converge any
more in the limit $n \to \infty$ when $x, y>1$. Instead, one has in
that case (see also Ref.~\cite{das})
\begin{eqnarray}
C_n(x,y) = 1 + \sum_{i=1}^{n}
i^{\tfrac{d-2}{2}} \; (xy)^i \propto \frac{n^{\tfrac{d-2}{2}} (x
  y)^{n+1}    }{x y -1} \; , \; x,y > 1 \;. \label{correl_das}
\end{eqnarray}
This leads to the expression for the density $\rho_{\infty}(x)$ for
$|x| > 1$ :
\begin{eqnarray}\label{density_kac_outer}
\rho_{\infty}(x) = \frac{1}{\pi (x^2-1)} \;,
\end{eqnarray}
independently of $d$. To understand better the divergence of $\rho_n(x)$ around $x = \pm 1$
(\ref{divergence_at_unity}) when $n \gg 1$, we focus on $\rho_n(x)$
{\it{around}} $x=1$. In the limit $n \gg 1$ and $1-x \ll 1$ keeping $y=n(1-x)$
fixed, one shows in Appendix \ref{appendix_density_kac} (see also~\cite{fyodorov}) that
\begin{eqnarray}
&&\rho_n(x) = n  \rho^K(n(1-x)) \;, \; \rho^K(y) = \frac{1}{\pi} \sqrt{ \frac{I_{d/2+1}(y)}{I_{d/2-1}(y)} -
  \left(\frac{I_{d/2}(y)}{I_{d/2-1}(y)} \right)^2 } \;, \label{scaling_density_kac}\\
&&I_m(y) = \int_{0}^{1} dx\; x^m \exp{(-2 y x)} \;. \label{def_Im}
\end{eqnarray}
One has $\rho^K(0) = \frac{2}{\pi} \frac{1}{d+2} \sqrt{\frac{d}{d+4}}$,
recovering the large $n$ behavior in  
Eq. (\ref{divergence_at_unity}) and its asymptotic behaviors are given
by~(see Appendix \ref{appendix_density_kac}) 
\begin{eqnarray} 
\rho^K(y) \sim 
\begin{cases}
\frac{1}{2 \pi y} \sqrt{\frac{d}{2}} \quad, \quad y \to + \infty
\label{density_large_uplus} \\
\frac{1}{2 \pi |y|} \quad, \quad y \to - \infty
\;. \label{density_large_uminus} 
\end{cases} 
\end{eqnarray}
For instance, one has
\begin{eqnarray}\label{scaled_density_kac_2d}
&&\rho^K(y) =  \frac{1}{2 \pi} \left(\frac{1}{y^2} -
\frac{1}{\sinh^2{y}}  \right)^{1/2} \quad, \quad {\rm{in}} \quad d=2 \;.
\end{eqnarray}
In the right panel of Fig. \ref{Fig1}, we show a plot of $n^{-1} \rho_n(x)$, where $\rho_n(x)$ is given in Eq. (\ref{density_kac_app}), 
as a function of $n(1-x)$ for $d=2$ and different large values
of $n=500, 1000$ together with the asymptotic results in
Eq. (\ref{scaled_density_kac_2d}) : we find a very good agreement with these 
analytic predictions (\ref{scaling_density_kac}, \ref{scaled_density_kac_2d}).

Having computed the mean density of real roots, we now focus on $\langle N_n
([a,b])\rangle$. On the interval $[0,1]$ the main contribution to the mean number of real roots comes, for large $n$, 
from the vicinity of $x=1$. Therefore, to compute $\langle N_n
[0,1]\rangle$ to leading order in $n$, one uses the scaling form for
the density in Eq.~(\ref{scaling_density_kac}), valid close to $x=1$,
and the asymptotic behavior in Eq.~(\ref{density_large_uplus}) to
obtain for $n \gg 1$ 
\begin{eqnarray}\label{number_kac_inner}
\langle N_n
[0,1]\rangle = \langle N_n [-1,0]\rangle = \int_0^n \rho^K(y) dy + {\cal O}(1) =
\frac{1}{2\pi} \sqrt{\frac{d}{2}} \log{n} + {\cal O}(1)  \;,
\end{eqnarray}
where the corrections of order ${\cal O}(1)$ receive contributions
from the whole interval $[0,1]$ (not only from the vicinity of $x =
\pm 1$).  

Similarly, one gets from Eq. (\ref{scaling_density_kac}) and Eq. (\ref{density_large_uminus}):
\begin{eqnarray}\label{number_kac_outer}
\langle N_n
[-\infty,-1]\rangle = \langle N_n [1,+\infty]\rangle = \int_0^n \rho^K(-y) dy + {\cal
  O}(1) = 
\frac{1}{2\pi} \log{n} + {\cal O}(1) \;,
\end{eqnarray}
which is independent of $d$ \cite{das}. From
Eqs~(\ref{number_kac_inner},~\ref{number_kac_outer})  
we compute the total number of roots on the real axis~:
\begin{eqnarray}
\langle  N_n( [-\infty,+\infty]) \rangle = \frac{1}{\pi} \left(1+
  \sqrt{\frac{d}{2}} \right) \log{n} + {\cal O}(1) \;,
\label{total_number_kac} 
\end{eqnarray} 
thus recovering, in a way similar to the one used in Ref.~\cite{edelman} for $d=2$, the result of \cite{das}. 
Notice that for $d=2$ the higher order terms of the large $n$ expansion in this formula
(\ref{total_number_kac}) have been obtained by various authors
(see for instance Ref.~\cite{edelman, wilkins}),
although, to our knowledge, they have not been computed for $d \neq 2$. 

In view of future purposes, we also
compute $\langle N_n [0,x] \rangle$ in the asymptotic limit where $n \gg 1$, and $0 <
1-x \ll 1$ with $n(1-x)$ kept fixed : 
\begin{eqnarray}
&&\langle N_n [0,x] \rangle = \langle N_n [0,1] \rangle - \eta_-(n(1-x)) \;,
\label{n_minus_kac} \\
&&\eta_-(y) = \int_{0}^{y} du \; \rho^K(u) \;, 
\end{eqnarray} 
such that $\eta_-(0)=0$ and with the asymptotic behavior obtained from Eq. (\ref{density_large_uplus})
\begin{eqnarray}
\eta_-(y) \sim \frac{1}{2 \pi} \sqrt{\frac{d}{2}} \log{y} \quad, \quad y \to
\infty \;.\label{eta_moins_large}
\end{eqnarray}

Similarly, we compute $\langle N_n [x,\infty] \rangle$ when $x >
    1$ and obtain for $n \gg 1$ and $0 < x-1 \ll 1$ keeping $y=n(x-1)$ fixed 
\begin{eqnarray}
&&\langle N_n [x,\infty] \rangle = \langle N_n [1,\infty]
\rangle - \eta_+(n(x-1)) \;, \label{n_plus_kac} \\
&&\eta_+(y) = \int_0^y du \rho^K(-u) \;,
\end{eqnarray}
such that $\eta_+(0) = 0$ and with the asymptotic behavior obtained from Eq. (\ref{density_large_uminus})
\begin{eqnarray}
\eta_+(y) \sim \frac{1}{2 \pi} \log{y} \quad, \quad y \to \infty \;.
\end{eqnarray}

We conclude this subsection by noting that, for $d=2$, the statistics of real roots of $K_n(x)$ is identical in the $4$
sub-intervals $[-\infty, -1], [-1,0], [0,1]$ and
$[1,+\infty]$. Instead, for $d \neq 2$,  the statistical behavior
of real roots of $K_n(x)$ depend on $d$ in the two inner intervals,
while it is identical to the case $d=2$ in the two outer
ones. In addition, we will see below that the polynomials $K_n(x)$~(\ref{def_kac_poly}) take independent values in these 4 subintervals.

\subsubsection{Weyl polynomials}

\begin{figure}[h]
%\begin{minipage}{0.5\linewidth}
\includegraphics[angle=-90,width=0.5\linewidth]{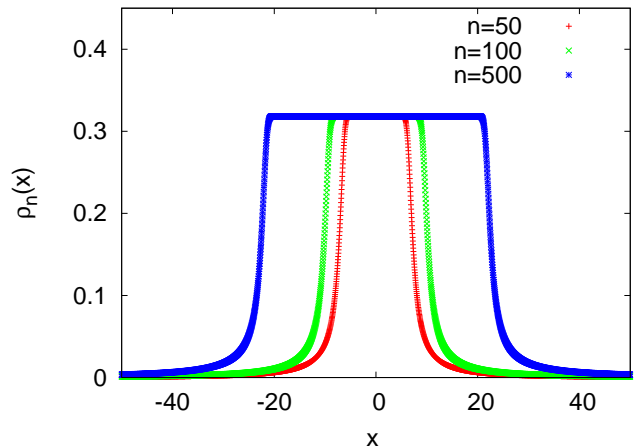}
%\end{minipage}\hfill
%\begin{minipage}{0.5\linewidth}
%\includegraphics[angle=-90,width=\linewidth]{scaled_density_weyl.eps}
%\end{minipage}
\caption{Mean density of real roots $\rho_n(x)$ given in Eq. (\ref{density_weyl}) for Weyl polynomials $W_n(x)$
  (\ref{def_weyl_poly}) as a function of $x$ for different values of
  $n = 50, 100$ and $500$.}\label{Fig2} 
\end{figure}

For Weyl polynomials $W_n(x)$ in Eq. (\ref{def_weyl_poly}), the
expression of the correlation function (\ref{correl_weyl}) together
with the expression for the density in Eq. (\ref{ek_formula}) yields
\begin{eqnarray}\label{density_weyl}
\rho_n(x) = \frac{1}{\pi} \sqrt{1 +
  \frac{x^{2n}(x^2-n-1)}{e^{x^2}\Gamma(n+1,x^2)} - \frac{x^{4n+2}
  }{[e^{x^2} \Gamma(n+1,x^2)]^2} } \;,
\end{eqnarray}
where $\Gamma(n,x) = \int_x^\infty dt e^{-t} t^{n-1}$ is the
incomplete gamma function \cite{grad}. In Fig. \ref{Fig2}, we show a
plot of $\rho_n(x)$ (\ref{density_weyl})  
for different values of $n=50, 100$ and $500$. One obtains straightforwardly, in the limit $n
\to \infty$ the uniform density
\begin{eqnarray}
\rho_\infty(x) = \frac{1}{\pi} \;. \label{mean_density_weyl}
\end{eqnarray}
For $n$ large but finite, the density is uniform like in
Eq. (\ref{mean_density_weyl}) up to $|x| \sim 
\sqrt{n}$ above which it vanishes (see Fig. \ref{Fig2}). Indeed, one shows in Appendix
\ref{appendix_density_weyl} that for $n 
\gg 1$, one has 
\begin{eqnarray}\label{density_scaling_weyl}
\rho_n(x) \sim 
\begin{cases}
\pi^{-1} \quad, \quad |x| \ll \sqrt{n} \;,\\
\tfrac{\sqrt{n}}{\pi x^2} \quad, \quad |x| \gg \sqrt{n} \;.
\end{cases}
\end{eqnarray}
One notices that this behavior of the density of real roots for Weyl polynomials (\ref{density_scaling_weyl}) is similar to the density 
of real eigenvalues for Ginibre random matrices \cite{edelman_ginibre}, {\it i.e.} random $n \times n$ matices formed from i.i.d. Gaussian entries.
Besides, from this scaling form (\ref{density_scaling_weyl}) one obtains the 
number of real roots in the interval $[-x,x]$, $x>0$, in the large $n$ limit as
\begin{eqnarray}
&&\langle  N_n [-x,x] \rangle = \int_{-x}^x dt \rho_n(t) \sim
\begin{cases}
2x/\pi \; ,\; x < \sqrt{n} \\
2\sqrt{n}/\pi \; ,\;x \geq \sqrt{n} \;,
\end{cases} \label{number_weyl}
\end{eqnarray}
from which one gets the total number of real roots for $n \gg 1$ (see also Ref.~\cite{leboeuf})
\begin{eqnarray}\label{total_weyl}
\langle  N_n [-\infty,+\infty] \rangle \sim \frac{2}{\pi} \, \sqrt{n} \;.
\end{eqnarray}
To our knowledge, the higher order terms in this large $n$ expansion
are not known.  

\subsubsection{Binomial polynomials}

For binomial polynomials, the computation of the density $\rho_n(x)$ is
straightforward. Indeed, using Eq. (\ref{correl_binom}) together with the
formula for the density (\ref{ek_formula}), one obtains
\begin{eqnarray}
\rho_n(x) = \sqrt{n} \rho^B(x) \quad, \quad \rho^B(x) =
\frac{1}{\pi(1+x^2)} \;, \label{mean_density_bp} 
\end{eqnarray}
exactly for
all $n > 1$ \cite{edelman, bleher}. 
\begin{figure}[h]
\includegraphics[angle=-90,width=0.5\linewidth]{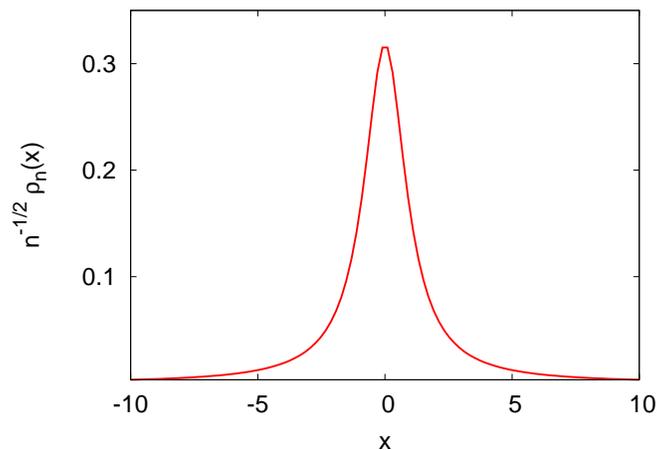}
\caption{Scaled mean density of real roots $n^{-1/2} \rho_n(x)$ as a
  function of $x$ (\ref{mean_density_bp}) for binomial polynomials
  $B_n(x)$ (\ref{def_binom_poly}).} \label{Fig3}
\end{figure}

In Fig. \ref{Fig3}, we show a plot of $\rho^B(x)$ as a function of $x$. This formula (\ref{mean_density_bp}) 
yields 
\begin{eqnarray}\label{number_binomial}
\langle  N_n[a,b] \rangle = \int_a^b  \, \rho_n(t) \,dt =
\frac{\sqrt{n}}{\pi} (\ArcTan{b} - \ArcTan{a}) \;, 
\end{eqnarray}
from which one gets the very simple result (see for instance
Ref.~\cite{edelman}) 
\begin{eqnarray}\label{total_bp}
\langle  N_n [-\infty,+\infty] \rangle = \sqrt{n} \;, \; {\rm exactly \;}
\; \forall n \;.
\end{eqnarray}

\subsection{"Gap probability'' on the real axis}

We now study another aspect of the statistical properties of the
real roots of these polynomials and focus on the probability
$P_0([a,b],n)$ that
they have no real root on a given interval $[a,b]$. The interval under
study will depend on the polynomials $K_n(x)$, $W_n(x)$ or
$B_n(x)$. 

\subsubsection{Results from the correlation function}

These polynomials, as a function of $x$, are Gaussian processes and
therefore their zero-crossing properties are completely determined by
the two-point correlators given in Eq. (\ref{correl_kac}-\ref{correl_binom}).

{\bf Generalized Kac polynomials}: For these polynomials $K_n(x)$, given the singularity
of the mean density $\rho_n(x)$ around $x = \pm 1$ (see Fig. \ref{Fig1}), it is natural to study separately
$P_0([0,x],n)$, for $x < 1$, and $P_0([x,\infty],n)$ for $x>1$. We first
    focus on $P_0([0,x],n)$ and reparametrize $K_n(x)$ with
a change of variable, $x = 1-1/t$.  One 
finds that the relevant scaling limit of ${C}_n(t,t')$ is
obtained for $t,t',n \to \infty$ keeping $\tilde t = t/n$ and $\tilde t' =
t'/n$ fixed. In that scaling limit the
discrete sum in Eq. (\ref{correl_kac}) can be viewed as a Riemann sum
and one finds
\begin{eqnarray}
C_n(x,y) \propto n^{d/2} I_{d/2-1}\left(\frac{1}{2\tilde t} +
\frac{1}{2 \tilde t'}\right) \;,
\end{eqnarray}
where $I_m(y)$ is defined in Eq. (\ref{def_Im}). Thus the normalized correlator ${\hat C}_n(t,t') \to {\cal
  C}(\tilde t, \tilde t')$ with the asymptotic behaviors (see
  Eq. (\ref{ik_large_uplus}))   
\begin{eqnarray} \label{correlator_asympt}
{\cal C}(\tilde t,\tilde t') \sim 
\begin{cases}
\left(4 \frac{{\tilde t \tilde t'}}{(\tilde t+\tilde
  t')^2}\right)^{\tfrac{d}{4}} \,,& \tilde t, \tilde 
t' \ll 1 \\ 
1 \,, & \tilde t,\tilde t' \gg 1 \;.
\end{cases}
\end{eqnarray}
Thus this correlator is exactly the same as the one
found for diffusion, ${\cal C}(\tilde t, \tilde t') = a(\tilde t,\tilde t')$ in Eq. (\ref{attprime}).
Since a Gaussian process is completely characterized by its two-point
correlator, we conclude that the diffusion process and the random polynomial 
are essentially the same Gaussian process and hence have the same
zero crossing properties. Therefore, in complete analogy 
with Eq. (\ref{fss1}) 
we propose the scaling form for generalized Kac polynomials  
\begin{eqnarray}
  P_0([0,x],n) = {\cal A}^-_{d,n} n^{-\theta(d)} h^-(n (1-x))
  \label{persist_poly1} \;, 
\end{eqnarray}   
where ${\cal A}^-_{d,n}$, which is independent of $x$, is such that $\lim_{n\to \infty}
\log{{\cal A}^-_{d,n}}/\log n = 0$ and $h^-(y) \to 1$ for $y \ll 1$ whereas $h^-(y) \sim
y^{\theta(d)}$ for $y \gg 1$, where 
$\theta(d)$ is the persistence exponent associated to the diffusion 
equation in dimension $d$. Defined in this way (\ref{persist_poly1}), $h^-(u)$ is a universal function (see below), although
the amplitude ${\cal A}^-_{d,n}$ is not. Note that $n$ here plays the role of
$L^2$ in diffusion problem while the variable $1-x$ is the 
analogue of the inverse time $1/t$. 

Similarly, we focus on $P_0([x,\infty],n)$, $x>1$, and reparametrize the
    polynomial with 
a change of variable, $x = 1+1/t$.  One 
finds that the relevant scaling limit of ${C}_n(t,t')$ is
obtained for $t,t',n \to \infty$ keeping $\tilde t = t/n$ and $\tilde t' =
t'/n$ fixed and one obtains
\begin{eqnarray}
C_n(x,y) \propto n^{d/2} I_{d/2-1}\left(-\frac{1}{2\tilde t} -
\frac{1}{2 \tilde t'}\right) \;,
\end{eqnarray}
where $I_m(y)$ is defined in Eq. (\ref{def_Im}). Thus ${\hat C}_n(x,y)
  \to {\cal C}(\tilde t, \tilde t')$ with the asymptotic behaviors (see
  Eq. (\ref{ik_large_uplus}))   
\begin{eqnarray} \label{correlator_asympt_bis}
{\cal C}(\tilde t,\tilde t') \sim 
\begin{cases}
\left(4 \frac{{\tilde t \tilde t'}}{(\tilde t+\tilde
  t')^2}\right)^{\tfrac{1}{2}} \,,& \tilde t, \tilde 
t' \ll 1 \\ 
1 \,, & \tilde t,\tilde t' \gg 1 \;, 
\end{cases}
\end{eqnarray}
independently of $d$. Therefore, in complete analogy 
with Eq. (\ref{fss1}) 
we propose the scaling form for random polynomials 
\begin{eqnarray}
P_0([x,\infty],n) = {\cal A}^+_{d,n} n^{-\theta(2)} h^+(n (x-1)) \;,
  \label{persist_poly_bis} 
\end{eqnarray}   
where ${\cal A}^+_{d,n}$, which is independent of $x$ is such that $\lim_{n\to \infty}
\log{{\cal A}^+_{d,n}}/\log n = 0$ and $h^+(y) \to 1$ for $y \ll 1$ whereas $h^-(y) \sim
y^{\theta(2)}$ for $y \gg 1$. 

Using the correlator $\hat C_{n}(x,y)$, one shows the statistical independence of the real roots of 
$K_n(x)$ in the four sub-intervals
$[-\infty, -1]$, $[-1,0]$, $[0,1]$ and $[1, +\infty]$. Consider for instance the intervals $[0,1]$ and $[1, +\infty]$. Given that the real
roots in the interval $[0, +\infty]$ are concentrated, for $n \gg 1$, around $x=1$ we introduce $x = 1-1/t$ and $y=1+1/t'$ and consider the
limit $t,t',n \to \infty$. One easily obtains
\beq\label{indep}
\hat C_{n}(1-1/t,1+1/t') \propto e^{-n/{\rm Max}(t,t')} \;,
\eeq 
which decays to $0$ exponentially for large $n$. Therefore one concludes that the zeros of $K_n(x)$ in the sub-intervals $[0,1]$ and $[1,\infty]$ are essentially independent. In a similar way, one shows that the real roots of $K_n(x)$ on the 
four subintervals delimited by $\pm 1$ are statistically independent.  Finally combining Eqs
(\ref{persist_poly1}, \ref{persist_poly_bis}) together with (\ref{indep}) one obtains the exact
asymptotic result for the probability of no real root as
\begin{eqnarray}\label{exact_no_root_kac}
P_0([-\infty,\infty],n) \propto n^{-2(\theta(d)+\theta(2))} \;.
\end{eqnarray}

We conclude this paragraph by presenting a heuristic argument which
allows to connect the zero crossing properties of the diffusion
equation to the one of the real roots of $K_n(x)$. For that purpose, we
consider the solution of the diffusion equation with random initial
condition (\ref{sol_eqdiff}) and we focus on $\phi(0,t)$, without any loss of generality. Following
Ref.~\cite{hilhorst}, one 
observes that the solid angle integration in that expression can be
absorbed into a redefinition of the random field, yielding
\begin{eqnarray}\label{sol_diff_heuristic}
\phi(0,t) = \frac{S_d^{1/2}}{(4 \pi t)^{d/2}} \int_0^L dr \;  r^{(d-1)/2}
e^{-r^2/t} \Psi(r) \;,\label{formule_hilhorst}
\end{eqnarray}
where $S_d$ is the surface of the $d$-dimensional unit sphere and
$\Psi(r)$ is given by \cite{hilhorst}
\beq
\Psi(r) = S_d^{-1/2} r^{-\tfrac{1}{2}(d-1)} \lim_{\Delta r \to 0} \frac{1}{\Delta r} \int_{r<x<r+\Delta r} d {\mathbf x} \; \psi (\mathbf x) \;, 
\eeq 
which is thus a random Gaussian variable of zero mean and
correlations $\langle \Psi(r) \Psi(r') \rangle = \delta(r-r')$. Performing the
change of variable $u = r^2$ in Eq. (\ref{formule_hilhorst}), one
obtains
\begin{eqnarray}
\phi(0,t) \propto \int_0^{L^2} du \; u^{\tfrac{d-2}{4}} e^{-u/t}
\tilde \Psi(u) \;, \; \langle \tilde \Psi(u) \tilde \Psi(u') \rangle = \delta(u-u') \;. \label{heuristic_diff}
\end{eqnarray}

On the other hand, if we focus on the real zeros of $K_n(x)$ in the
interval $[0,x]$ with $x < 1$, we know that these zeros accumulate
in the vicinity of $x=1$. Therefore, in terms of $x = 1-1/t$ one has
\begin{eqnarray}\label{k_heuristic}
K_n(x) \sim a_0 + \sum_{i=1}^n i^{(d-2)/4}e^{-i/t} a_i \;.
\end{eqnarray} 
By approximating the discrete sum in the above
expression~(\ref{k_heuristic}) by an integral, one sees
that $K_n(x)$ is similar to the solution of the diffusion equation in
Eq. (\ref{heuristic_diff}) where $L^2$ 
is replaced by $n$ and $1-x$ by $1/t$. Therefore one understands qualitatively why the zero crossing properties 
of these two processes coincide.

{\bf Weyl polynomials:} To analyse the correlation function in Eq. (\ref{correl_weyl}) in
the large $n$ limit we write it as
\begin{eqnarray}\label{correl_weyl_gamma}
C_n (x,y) &=& \langle W_n(x) W_n(y) \rangle = \sum_{i=0}^n \frac{(xy)^i}{i
  !} = e^{xy} \frac{\Gamma(n+1,xy)}{\Gamma(n+1)} \;,
\end{eqnarray}
where the last equality can easily be obtained using the recursion relation $\Gamma(n+1,z) = n \Gamma(n,z)+ e^{-z} z^n$. The behavior of $\Gamma(n,z)$ for large $n$ is analysed in detail in Appendix \ref{appendix_density_weyl}. From the results obtained in Eq. (\ref{expansion_gamma_smallu}, \ref{expansion_gamma_largeu}), one sees that the correlation function
 $C_n (x,y)$ in Eq. (\ref{correl_weyl_gamma}) behaves differently for $x y < n$ and $x y >n$. 

For $x y < n$, Eq. (\ref{expansion_gamma_smallu}) shows that $\Gamma(n+1,x y) \to \Gamma(n+1)$ for large $n$ so that one finds 
that $\hat C_n (x,y) \to {\cal C}(x,y)$ with
\begin{eqnarray}\label{asympt_weyl_small_arg} 
{\cal C}(x,y) = \exp{[ -\tfrac{1}{2}(x- y)^2]} \;.
\end{eqnarray}
%\begin{eqnarray}
%{\cal C}_n(\tilde x,\tilde y) \sim 
%\begin{cases}
%e^{-\tfrac{n}{2}(\tilde x- \tilde y)^2} \quad \tilde x  \tilde y < 1
%\label{asympt_weyl_small_arg} \\
%\frac{\sqrt{\tilde x^2-1}\sqrt{\tilde y^2-1}}{\tilde x \tilde y-1} 
%\quad \tilde x \tilde y \geq 1 \;.\label{asympt_weyl_large_arg}
%\end{cases}
%\end{eqnarray}
Interestingly Eq. (\ref{asympt_weyl_small_arg}) shows that inside the interval $[-\sqrt{n}, \sqrt{n}]$, 
$W_n(x)$ is exactly the GSP
characterizing the zero crossing properties of the diffusion field in 
the limit of infinite dimension $d\to \infty$
(\ref{correlator_larged}). Therefore one expects $P_0([-x,x],n)$, the
probability that $W_n(x)$ has no real root in the interval $[-x,x]$,
with $1 \ll x \leq \sqrt{n}$, to behave as 
\begin{eqnarray}\label{persist_weyl_1}
P_0([-x,x],n) \propto \exp{(-2 \theta_\infty x)} \;.
\end{eqnarray}
For $x y > n$, the behavior of $C_n(x,y)$ is quite different. Indeed, using the asymptotic behavior in
Eq. (\ref{expansion_gamma_largeu}), one shows that the relevant scaling limit is obtained for $x,y,n \to \infty$ keeping $\tilde x = x/\sqrt{n}$ and
$\tilde y = y/\sqrt{n}$ fixed such that $\hat C_n(x,y) \to {\cal C}(\tilde x,\tilde y)$ with
\begin{eqnarray}
{\cal C}(\tilde x,\tilde y) = \frac{\sqrt{\tilde x^2-1}\sqrt{\tilde
    y^2-1}}{\tilde x \tilde y-1}  \quad , \; \tilde x \tilde y \geq 1
\;,\label{asympt_weyl_large_arg} \;.
\end{eqnarray}
Performing the change of variable $x \to \tilde x = \sqrt{n}+1/(\sqrt{n} \tilde t)$ one easily obtains that ${\cal C}(\tilde t,
\tilde t')$ behaves like in
Eq. (\ref{correlator_asympt_bis}). Therefore, by analogy with
Eq. (\ref{persist_poly_bis}), one deduces that, for $0< x - \sqrt{n} \ll 1$, $n \gg 1$ keeping
$\sqrt{n} (x-\sqrt{n})$ fixed, one has
\begin{eqnarray}\label{persist_weyl_2}
P_0([x,\infty],n) \propto n^{-\tfrac{\theta(2)}{2}} 
    w(\sqrt{n}(x-\sqrt{n}))  \;,
\end{eqnarray}
where $w(u) \sim c^{\rm st }$ for $u \ll 1$ and $w(u) \propto
u^{\theta(2)}$ for $u \gg 1$. In addition, following the arguments presented above (see Eq. (\ref{indep})),
one shows that these two outer-intervals $[-\infty, -\sqrt{n}]$ and $[\sqrt{n},+\infty]$ are
statistically independent, such that
\bea\label{persist_weyl_3}
P_0([-\infty, -\sqrt{n}] \cup [\sqrt{n},+\infty],n) \propto n^{-\theta(2)} \;.
\eea
However, given the behavior of the correlator ${\cal C}(x,y)$ for $x y > n$ in Eq. (\ref{asympt_weyl_small_arg}) the inner and outer intervals are not independent and the probability of 
no root on the real axis is not the product of the probabilities in 
Eq. (\ref{persist_weyl_1}) evaluated in $x=\sqrt{n}$ and the the one
in Eq. (\ref{persist_weyl_3}) : the effect of these correlations will
be discussed below. 

{\bf Binomial polynomials:} In that case, one can extract information directly
from the correlation function in Eq.~(\ref{correl_binom}) by focusing in the
limit $x, y \to 0$. In that limit the normalized correlation function $\hat
C_n(x,y)$ is given by
\bea
\hat C_n(x,y) = \frac{(1+x\,y)^n}{[(1+x^2)(1+y^2)]^{\tfrac{n}{2}}} \sim \exp{\left[-\tfrac{n}{2}(x-y)^2\right]} \;, \; x,y \ll 1 \;.
\eea
Thus in the large $n$ limit, the probability $P_0([a,b],n)$ that
binomial polynomials have no real root in the interval $[a,b]$ with
$a<b \ll 1$, $n^{-\tfrac{1}{2}} \ll b-a$ 
behaves like 
\begin{eqnarray}\label{persist_asympt_bp}
P_0([a,b],n) \propto \exp{\left[-\theta_\infty \sqrt{n} (b-a)\right]}
\;, \;  a<b \ll 1 \;, \; n^{-\tfrac{1}{2}} \ll b-a  \;, 
\end{eqnarray} 
which is an exact statement. However it
is a more difficult task to obtain the behavior of
$P_0([a,b],n)$ for an arbitrary interval $[a,b]$ and eventually obtain the
probability of no root on the entire real axis for this class of
polynomials (\ref{def_binom_poly})~: this will be achieved in the next
sections.  

To conclude this paragraph, we have shown that the analysis of the
correlation function $C_n(x,y)$ yields important {\it exact} results
for the gap probabilities. Indeed, for generalized Kac polynomials, we
obtained the important results in Eq. (\ref{persist_poly1}) and
Eq. (\ref{persist_poly_bis}) which yield the exact result in
Eq. (\ref{exact_no_root_kac}). For Weyl 
polynomials the study of the correlation function allowed us to obtain the results in Eq. (\ref{persist_weyl_1}) and Eq. (\ref{persist_weyl_2}). Finally, for
Binomial polynomials we obtained, from the correlation funtion, the asymptotic behavior  in Eq. (\ref{persist_asympt_bp}), which will be useful in the following.

\subsubsection{Mean-Field description : Poisson approximation} 

To calculate the gap probabilities and the associated scaling functions, we first develop a very simple mean field theory. This theory, albeit approximate as it neglects the correlations between zeros, is simple, intuitive and qualitatively correct. We will see later how one can improve systematically this mean field theory to get answers that are even quantitatively accurate. As a first step,
we neglect the correlations between the real roots and simply consider
that these roots are randomly and independently distributed on the real
axis with some local density $\rho_n(x)$ at point $x$. Within this approximation the probability $P_k([a,b],n)$ that these polynomials have exactly $k$ real roots satisfies the equation 
\beq\label{recurrence_mf}
\frac{\partial P_{k+1}([a,b],n)}{\partial b} = \rho_n(b) [ P_{k}([a,b],n)-P_{k+1}([a,b],n) ] \;,
\eeq
together with the normalization condition $\sum_{k\geq0} P_{k}([a,b],n) = 1$ and $P_{k}([a,b],n) = \delta_{k,0}$ when $a=b$. In the large $n$ limit (where one can omit the constraint $P_{k>n}([a,b],n) = 0$), $P_{k}([a,b],n)$ is given by a non-homogeneous Poisson distribution
\beq\label{poissonian_1}
P_{k}([a,b],n) = \frac{\mu^k}{k!} \,e^{-\mu} \;, \; \mu = \langle N_n[a,b] \rangle = \int_a^b \rho_n(x) dx \;,
\eeq
which clearly satisfies Eq. (\ref{recurrence_mf}). In particular, this mean field approximation (\ref{poissonian_1}) yields the gap probability 
\begin{eqnarray}
P_0([a,b],n) = \exp{\left(-\langle {N}_n[a,b] \rangle \right)} =
  \exp{\left(-\int_a^b 
  \rho_n(x) \; dx\right)} \;. \label{poissonian_gen} 
\end{eqnarray}  

When applied to Generalized Kac polynomials  $K_n(x)$, for which we
obtained $\langle {N}_n[0,x] \rangle $ in Eq. (\ref{n_minus_kac}), this
mean-field approximation (\ref{poissonian_gen}) yields in the scaling
limit $n\to \infty$, $1-x \to 0$, $n(1-x) > 0$ fixed
\begin{eqnarray}
P_0([0,x],n) = {\cal A}_{d,n}^- n^{-\tfrac{1}{2\pi} \sqrt{\tfrac{d}{2}}}
\exp{\left[\eta_-(n(1-x)) \right]} \;,
\end{eqnarray}
where, from Eq.~(\ref{number_kac_inner}), $\log{\cal A}_{d,n}^- =
o(\log{n})$. This mean-field approximation thus yields the correct
scaling from for $P_0([0,x],n)$ as in
Eq. (\ref{persist_poly1}), with the non trivial predictions for the exponent and scaling function
\begin{eqnarray}\label{poissonian_kac}
\theta^{\rm MF}(d) = \frac{1}{2\pi} \sqrt{\frac{d}{2}} \hspace*{0.2cm}, \hspace*{0.2cm} h^-(u) =
\exp{\left(\int_0^u dy \rho^K(y) \right)} \;,
\end{eqnarray}
with the asymptotic behavior $h^-(u) \sim 1$ for $u \ll 1$
and, using the asymptotic behavior obtained in
Eq. (\ref{eta_moins_large}), $h^-(u)  \sim
u^{\theta^{\rm MF}(d)}$.

Similarly, this mean-field approximation applied to Weyl polynomials
$W_n(x)$, for which we obtained $\langle {N}_n[-x,x]\rangle $ in
Eq. (\ref{number_weyl}), yields the scaling form
\begin{eqnarray}
P_0([-x,x],n)   \sim
\begin{cases}
\exp{(-2 x/\pi)} \;, \; x < \sqrt{n} \\
\exp{(-2 \sqrt{n}/\pi)} \;, \; x \geq \sqrt{n} \;.
\end{cases} \label{weyl_poissonian}
\end{eqnarray}
Notice that this mean-field approximation gives an approximation of
$\theta^{\rm MF}_\infty = \pi^{-1} = 0.31831...$ (see Eq. (\ref{persist_weyl_1})), which is consistent, using the relation $\theta(d) = 2^{-3/2} \theta_\infty \sqrt{d} $ for large $d$, with Eq. (\ref{poissonian_kac}). 

Finally, if one uses this mean-field approximation to study binomial
polynomials $B_n(x)$, one obains, using the expression $\langle {N}_n[a,b] \rangle$ 
given in Eq. (\ref{number_binomial})
\begin{eqnarray} 
P_0([a,b],n) = \exp{\left[-\frac{\sqrt{n}}{\pi}
  (\ArcTan{a} - \ArcTan{b})\right]} \;,
\end{eqnarray}
which again, according to Eq. (\ref{persist_asympt_bp}), gives the mean-field approximation for
$\theta^{\rm MF}_\infty = \pi^{-1}$, as above.

\subsubsection{Beyond Mean-Field : a systematic approach}

We will now show that this mean field approximation (\ref{poissonian_gen}) can actually be
improved systematically. For that purpose, one considers the probability
$P_k([a,b],n)$ that such polynomials as in Eq. (\ref{def_kac_poly},
\ref{def_weyl_poly}, \ref{def_binom_poly}) have exactly $k$ 
real roots in the interval $[a,b]$. Following
Ref.~\cite{satya_partial}, one introduces the generating function
\begin{eqnarray}\label{generating_function}
\hat P_n(p,[a,b]) = \sum_{k=0}^\infty p^k P_k([a,b],n) \;,
\end{eqnarray}
where $\hat P_n(p,[a,b])$ can be interpreted as a persistence probability with partial survival \cite{satya_partial}.
For a smooth process, it turns out that $\hat \theta_n(p,[a,b]) = -\log{(\hat P_n(p,[a,b]))}$ depends {\it continuously} on
$p$ : this was shown exactly for the random acceleration process (see Eq. (\ref{def_RAP}) below) and approximately using
the IIA - and further checked numerically - for the diffusion equation with random initial conditions \cite{satya_partial}.
Thus one has
\begin{eqnarray}
\hat \theta_n(p,[a,b]) = -\log{(\hat P_n(p,[a,b]))} = -\sum_{r=1}^\infty
  \frac{(\log{(p)})^r}{r !} 
  \langle N^r_n([a,b])  \rangle_c  \;,\label{exp_hat_theta}
\end{eqnarray}
where the notation $\langle...\rangle_c$ stands for a connected
average. Here we are interested in $\hat P_n(p=0,[a,b]) = P_0([a,b],n)$ and
the idea, given that $\hat \theta_n(p=1,[a,b]) = 0$ is to expand $\hat \theta_n(p,[a,b])$
around $p=1$ in an $\epsilon$-expansion with $p=1-\epsilon$. This yields 
\begin{eqnarray}
\hat \theta_n(1-\epsilon,[a,b]) = \sum_{r=1}^\infty {\mathfrak a}_{r,n}([a,b]) \epsilon^r \;,
\label{epsilon_expansion} 
\end{eqnarray}
where ${\mathfrak a}_{r,n}([a,b])$ are linear combinations of the cumulants 
$\langle N^m_n[a,b]  \rangle_c$, with $m \leq r$. For instance,
\begin{eqnarray}
&&{\mathfrak a}_{1,n}([a,b]) = \langle N_n[a,b] \rangle  \;, \;
{\mathfrak a}_{2,n}([a,b]) = \tfrac{1}{2}\left(\langle N_n[a,b] \rangle- \langle
N^2_n[a,b] \rangle_c   \right) \;,   \label{explicit_ar} \\
&&{\mathfrak a}_{3,n}([a,b]) = \left(\frac{\langle N_n[a,b] \rangle}{3} - \frac{\langle
N^2_n[a,b] \rangle_c}{2} + \frac{\langle
N^3_n[a,b] \rangle_c}{6}   \right) \;. \nonumber
\end{eqnarray}
Thus one sees that if one restricts the $\epsilon$-expansion in
Eq. (\ref{epsilon_expansion}) to first order, and set $\epsilon=1$,
one recovers the mean-field approximation (\ref{poissonian_gen}),
using ${\mathfrak a}_{1,n}([a,b]) = \langle {N}_n[a,b] \rangle$. Higher order terms in this
$\epsilon$-expansion allow to improve systematically this mean-field approach.

{\bf Kac polynomials for $d=2$:} We first illustrate this $\epsilon$
expansion 
for Kac polynomials $K_n(x)$ for $d=2$ where we compute $\hat
\theta_n(1-\epsilon,[0,x])$ up to order ${\cal O}(\epsilon^2)$. In that
purpose, we compute ${\mathfrak a}_{2,n}([0,x])$. In Appendix
\ref{appendix_scaling_kac}, we show 
that in the scaling limit $(1-x) \to 0$, and $n \to \infty$ keeping
the product $n(1-x)$ fixed one has, similarly to the Eq. (\ref{n_minus_kac}) for the first moment,
\begin{eqnarray}
\langle N^2_n[0,x] \rangle_c = \langle N^2_n[0,1] \rangle_c - \nu_-(n(1-x)) 
\label{scaling_cumulant_kac} 
\end{eqnarray}  
where $\nu_-(y)$, given in Eq. (\ref{expr_g}), is such that $\nu_-(y)  \to 0$ for $y \ll 1$ and  
\bea
\nu_-(y) \sim \left(\frac{1}{\pi}-\frac{2}{\pi^2} \right)\log{y} \;, \; y \gg 1 \;.
\eea
Notice that
$\langle N^2_n[0,1] \rangle_c$ in Eq. (\ref{scaling_cumulant_kac}) has
been computed in Ref.~\cite{maslova}, yielding for large $n$
\beq\label{variance_maslova}
\langle N^2_n[0,1] \rangle_c = \left(\frac{1}{\pi} - \frac{2}{\pi^2}\right) \log{n} + o(\log{n}) \;,
\eeq
although higher order terms in this large $n$ expansion are not known. 
Combining Eq. (\ref{epsilon_expansion}, \ref{explicit_ar},
\ref{scaling_cumulant_kac}) together with the expression for $\nu_-(y)$ in
Eq. (\ref{expr_g}), one obtains in the scaling limit
\bea\label{order_2_kac}
&&\hat \theta_n(1-\epsilon,[0,x]) = \left(\epsilon
+\tfrac{\epsilon^2}{2}\right) \langle N_n([0,1]) \rangle -
\tfrac{\epsilon^2}{2}  \langle N_n([0,1])^2 \rangle_c \\
&& - \epsilon \int_0^y du \, \rho^K(u) + \epsilon^2 \int_0^y du_1
\int_{u_1}^\infty du_2 \left( \tilde {\cal K}(u_1,u_2) - 
\rho^K(u_1) \rho^K(u_2) \right) + {\cal O}(\epsilon^3) \;,\nonumber
\eea
with $y = n(1-x)$. In this Eq.~(\ref{order_2_kac}), $\rho^K(u)$ is
given in Eq.~(\ref{scaled_density_kac_2d}), and $\tilde {\cal
  K}(u_1,u_2)$, which we study in detail in Appendix~\ref{appendix_scaling_kac}, is essentially the two-point 
correlation function of real roots inside the peak of the density around $x = 1$ (see right panel of Fig. \ref{Fig1}). 
Setting $\epsilon = 1$ in this expression to order ${\cal O}(\epsilon^2)$ (\ref{order_2_kac}) one obtains, for $d=2$
\bea\label{order_2_kac_scaling}
&&P_0([0,x],n) = {\cal A}^-_{d,n} n^{-\theta(2)} h^-(n(x-1)) \; , \; \theta(2) = \frac{\pi + 4}{4 \pi^2} =0.180899... \;, \\
&&h^{-}(y) = \exp{\left(  \int_0^y du \, \rho^K(u) -  \int_0^y du_1
\int_{u_1}^\infty du_2 \left( \tilde {\cal K}(u_1,u_2) - \rho^K(u_1) \rho^K(u_2) \right) \right)  } \;,
\eea
with $\log{{\cal A}^-_{d,n}} = o(\log{n})$ (see Eqs (\ref{number_kac_inner}, \ref{variance_maslova})). Note that the value of the exponent $\theta(2)$ up to second order as given in this Eq.~(\ref{order_2_kac_scaling}) was computed in Ref. \cite{satya_partial}.
In the next sections, we will show, using numerics, that this second order calculation (\ref{order_2_kac_scaling}) is a true improvement 
upon the mean-field approximation (\ref{poissonian_kac}).

{\bf Weyl polynomials:} We now compute $\hat \theta_n(1-\epsilon,[-x,x])$, $x < \sqrt{n}$
for Weyl polynomials up to order ${\cal O}(\epsilon^2)$. In that
purpose we compute ${\mathfrak a}_{2,n}([-x,x])$. In Appendix
\ref{appendix_scaling_weyl}, one shows that for $x < \sqrt{n}$ fixed
and $n \gg 1$, one has  
\bea 
\langle N^2_n[-x,x]\rangle_c = \nu(x) \;, \label{scaling_cumulant_weyl}
\eea
with $\nu(x)$ given in Eq. (\ref{g_tilde_weyl}). Combining
Eq. (\ref{epsilon_expansion}, \ref{explicit_ar},
\ref{scaling_cumulant_weyl}) together with the expression for $\nu(x)$ in
Eq. (\ref{g_tilde_weyl}), one obtains for large $n$
\bea\label{order_2_weyl}
&&\hat \theta_n(1-\epsilon,[-x,x]) = \epsilon \frac{2 x}{\pi} + 2 \epsilon^2
\int_0^x ds (s-x) (\tilde {\cal W}(s)-\pi^{-2}) + {\cal O}(\epsilon^3) \;, 
\eea
where $\tilde {\cal W}(s)$, given in Eq. (\ref{tilde_w}), is the two point correlation function of real roots in the interval $[-\sqrt{n}, \sqrt{n}]$. 
Setting $\epsilon=1$ in this expression up to order ${\cal O}(\epsilon^2)$
(\ref{order_2_weyl}), one obtains, for $x<\sqrt{n}$ 
\bea
P_0([-x,x],n) = \exp{\left(-\frac{2x}{\pi} -\int_0^{2 x} ds (s-2x) (\tilde {\cal
    W}(s)-\pi^{-2})\right)}  \;.
\eea
Using Eq. (\ref{large_g_weyl}), one obtains its large $x$ behavior as
in Eq. (\ref{persist_weyl_1}) with the value of $\theta_\infty$ up to
order ${\cal O}(\epsilon^2)$ 
\beq\label{proba_second_order_weyl}
\theta_{\infty} = \frac{1}{\pi} - \int_0^\infty ds (\tilde {\cal
  W}(s)-\pi^{-2}) = 0.386471... \;,
\eeq 
which should be compared with the numerical value $\theta_{\infty,{\rm sim}} =0.417(3)$ \cite{newman_diffusion}.

If we focus instead on the outer intervals $[-\infty, -\sqrt{n}] \cup [\sqrt{n}, \infty]$, this $\epsilon$ expansion is essentially
similar to the one performed for Kac polynomials and $d=2$, given the behavior of the correlator in Eq. (\ref{asympt_weyl_large_arg}). More interestingly, if we are interested in the computation of the probability of no root
on the real axis, this $\epsilon$ expansion is able to take into account (perturbatively) the correlations between the 
inner and outer intervals, which, as discussed below
Eq. (\ref{persist_weyl_3}), can be seen in the correlation
function~(\ref{asympt_weyl_small_arg}). Doing so, one obtains 
that 
\beq\label{correl_weyl_inter}
P_0([-\infty, \infty],n) =      P_0([-\sqrt{n}, \sqrt{n}],n)   P_0([-\infty, -\sqrt{n}] \cup [\sqrt{n}, \infty],n) \mathfrak{p}_n \;,
\eeq
where the computation of  $\mathfrak{p}_n$ is similar to the one carried out in Appendix \ref{appendix_scaling_weyl} where the quantity ${\cal W}_n(t_1,t_2)$ in Eq.~(\ref{def_Wn}) involves $t_1 \in [-\sqrt{n}, \sqrt{n}]$ and $t_2 \in [-\infty, -\sqrt{n}] \cup [\sqrt{n}, \infty]$ such that $t_1 t_2 < 1$. The computation up to order ${\cal O}(\epsilon^2)$  shows that $\mathfrak{p}_n \propto n^{\tau}$ with $\tau > 0$. Therefore, on the basis of this result together with Eqs (\ref{persist_weyl_1}, \ref{persist_weyl_3}), one expects
\beq\label{persist_weyl_entire}
P_0([-\infty, \infty],n) \sim n^{-\gamma} \exp{(-2\theta_\infty \sqrt{n})} \;,
\eeq
where the exponent $\gamma$ is a priori unknown. Below, we will confront this statement with numerical simulations.  

{\bf Binomial polynomials:} We now focus on $\hat \theta_n(1-\epsilon,n)$ for binomial polynomials. In Ref.~\cite{bleher}, it was shown that for large $n$ and {\it all} $a,b > n^{-\tfrac{1}{2}}$
\beq\label{scaling_cumulant_bp_1}
\langle N^2_n[a,b]\rangle_c \propto \beta_2 \langle
N_n[a,b]\rangle \;,
\eeq 
where $\beta_2$ is a constant, independent of $n, a$ and $b$. It has an analytic expression in term of an integral involving elementary functions, with $\beta_2 = 0.571731...$. This expression (\ref{scaling_cumulant_bp_1}) yields in the large $n$ limit, 
${\mathfrak a}_{2,n}([a,b])~=~(1/2~-~\beta_2~)~\langle N_n[a,b]\rangle$ which together with the asymptotic behavior in 
Eq. (\ref{persist_asympt_bp}) and the cumulant expansion of Eq. (\ref{epsilon_expansion})
allows to compute  $\theta_\infty$ up to order ${\cal
  O}(\epsilon^2)$. We have checked that this coincides with the one
obtained in Eq. (\ref{proba_second_order_weyl}). More generally, one expects that for all integer $m > 0$
\beq\label{scaling_cumulant_bp_2}
\langle N^m_n[a,b]\rangle_c \propto \beta_m \langle N_n[a,b]\rangle \;,
\eeq
where $\beta_m$ is a constant, independent of $n, a$ and $b$ and in
Appendix \ref{appendix_scaling_bp} we explicitly show the mechanism
leading to 
this relation (\ref{scaling_cumulant_bp_2}) for $k=3$. From these relations
(\ref{scaling_cumulant_bp_2}) and the structure of the cumulant expansion in Eq. (\ref{epsilon_expansion}), one expects that
$P_0([a,b],n) \propto \exp{[-\omega \langle
  N_n[a,b]\rangle]}$, where $\omega$ is a linear combination of the coefficients $\beta_m$. Finally, this expression has to match the exact asymptotic 
behavior of $P_0([a,b],n)$ for $a<b \ll 1$ and $n^{-\tfrac{1}{2}} \ll b-a$ derived in
Eq.~(\ref{persist_asympt_bp}). Thus one has $\omega = \pi \theta_{\infty}$ 
so that one obtains the {\it exact} result 
\bea \label{persist_bp_exact}
P_0([a,b],n) \propto \exp{\left[-\sqrt{n}
  \theta_{\infty}(\ArcTan{b}-\ArcTan{a}) \right]} \;,
\eea
from which we obtain the exact expression for the probability of no real
root for $B_n(x)$ in the large $n$ limit as
\beq\label{persist_bp_exact_2}
P_0([-\infty,+\infty],n) \propto
    \exp{\left(-\sqrt{n} \pi \theta_{\infty} \right)} \;.
\eeq
Below, we check this analytical result (\ref{persist_bp_exact}) using
numerical simulations.

\subsubsection{Numerical results}

\begin{figure}[h]
\begin{minipage}{0.5\linewidth}
\includegraphics[angle=-90,width=\linewidth]{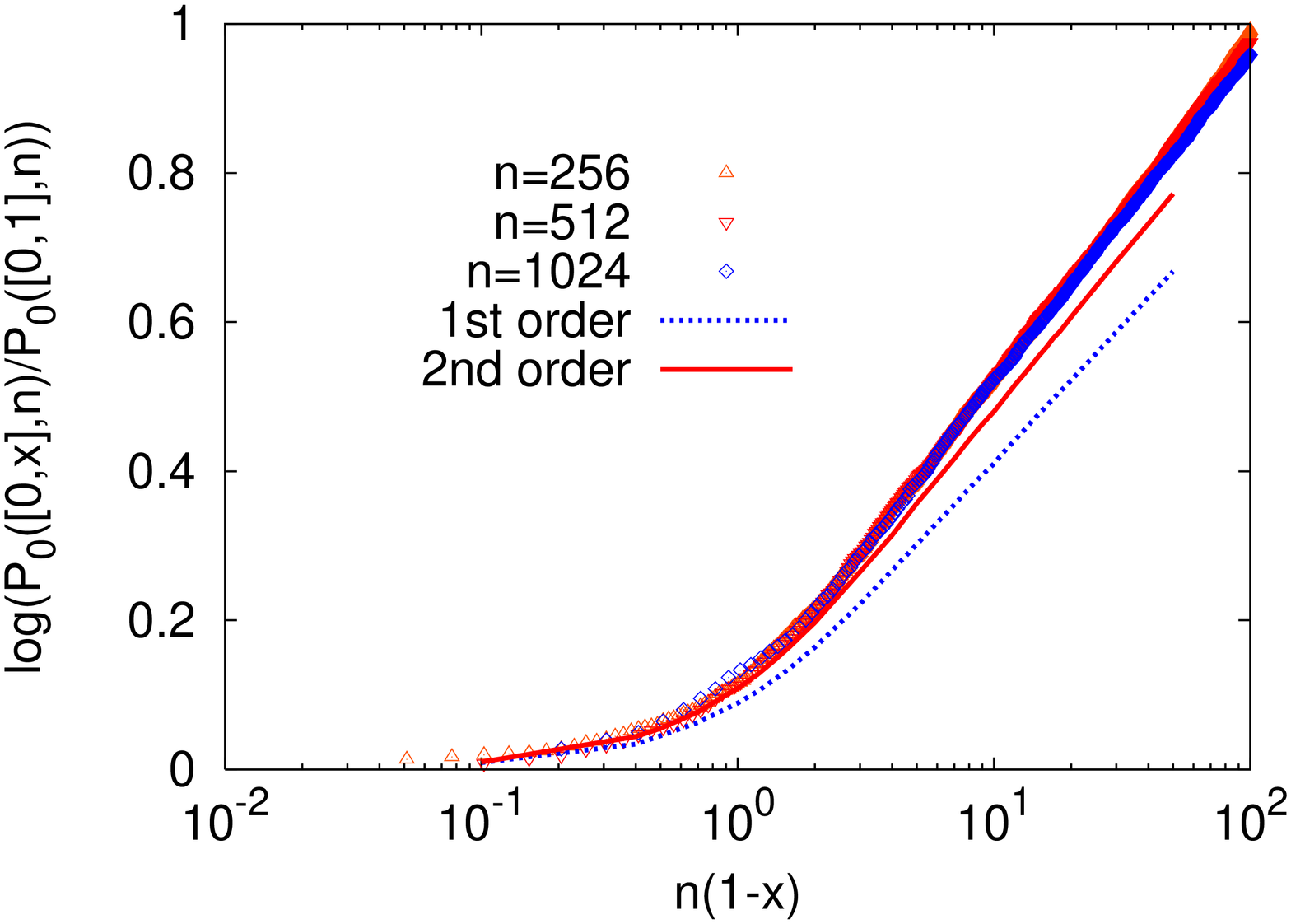}
\end{minipage}\hfill
\begin{minipage}{0.5\linewidth}
\includegraphics[angle=-90,width=\linewidth]{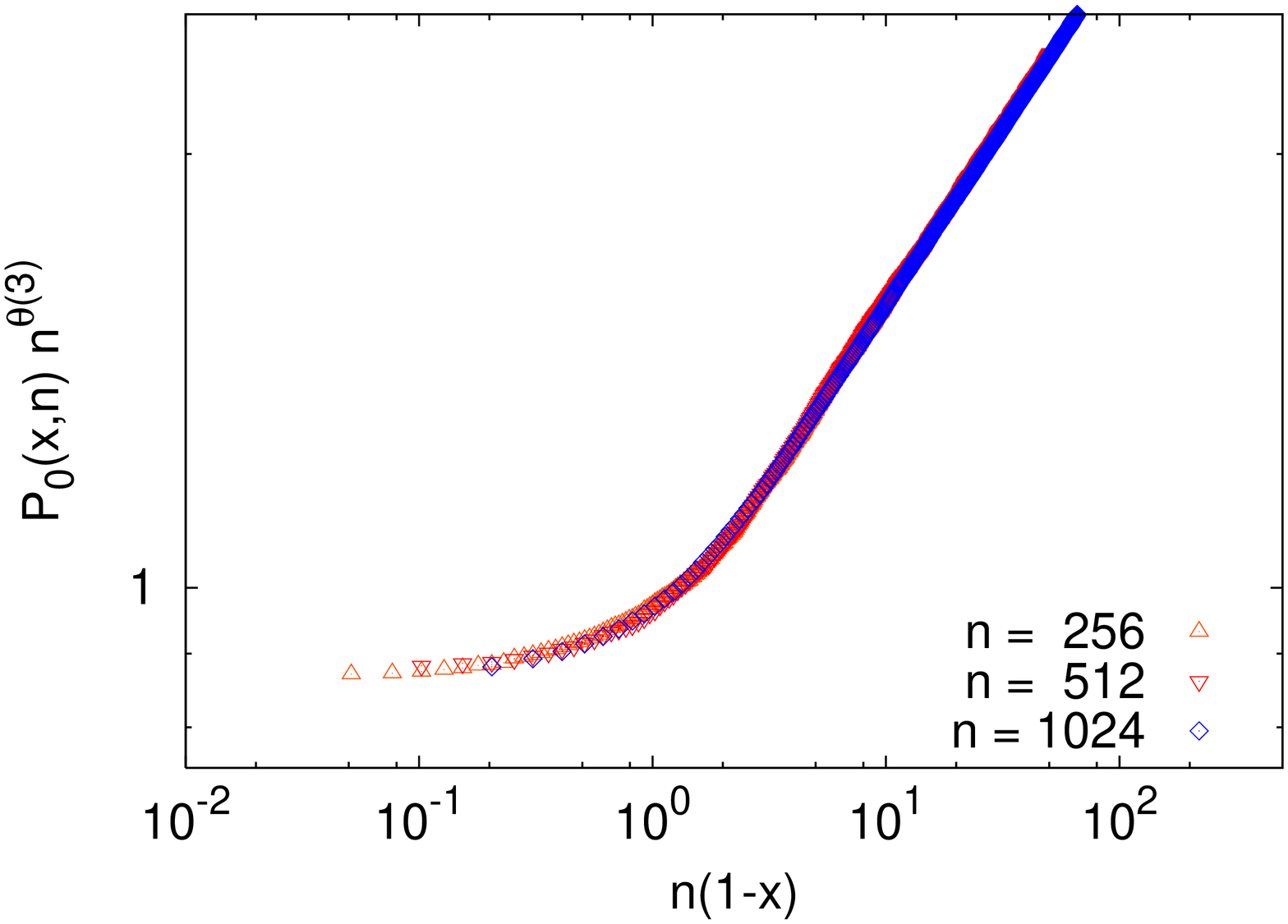}
\end{minipage}
\caption{{\bf Left} : Plot of $\log[P_0([0,x],n)/P_0([0,1],n)]$ for
  Kac polynomials $K_n(x)$ and $d=2$ as a function of $n(1-x)$ for different degrees $n=256, 512, 1024$.
The dotted is the result of the Mean Field prediction
  (\ref{poissonian_kac}) and the solid line the result of the second
  order calculation in the $\epsilon$ expansion
(\ref{order_2_kac}).  {\bf Right} : Plot of $n^{\theta(3)}
  P_0([0,x],n)$ on a log-log scale for Kac polynomials $K_n(x)$ and
  $d=3$ with $\theta(3) = 0.238(4)$ as a function 
  of $n(1-x)$ for different degrees $n=256, 512, 1024$.}\label{Fig4}
\end{figure}

{\bf Kac polynomials:} We first focus on the interval $[0,1]$ and check numerically the scaling forms for
$P_0([0,x],n)$ in Eq.~(\ref{fss1}) for $K_n(x)$ and for different
values of $d$. In 
each case, this probability is obtained by averaging over $10^4$
realizations of the random variables $a_i$'s in
Eq. (\ref{def_kac_poly}), drawn independently from a Gaussian
distribution of unit variance. In Ref.~\cite{us_prl}, we already presented numerical results for $P_0([0,x],n)$ in $d=2$ and
we also checked that $P_0([0,x],n) \propto n^{-\theta(2)}$ with
$\theta(2) = 0.187(1)$.  In the  
left panel of Fig. \ref{Fig4}, we show a plot of $\log[P_0([0,x],n)/P_0([0,1],n)]$ for $d=2$
as a function of the scaled variable $n(1-x)$ for different values of $n$. According to Eq. (\ref{fss1}), together with 
the good collapse of the curves for different values of $n$, this allows for a numerical computation of the scaling function 
$h^{-}(y)$. We have checked that different distributions of the random coefficients either $a_i = \pm 1$ or rectangular distribution
yield the same scaling function $h^{-}(y)$, suggesting that this function is indeed universal. On the same figure, the dotted line is
the result of  Mean Field approximation (\ref{poissonian_kac}), or first order in the $\epsilon$ expansion, 
and the solid line is the analytical result of the second order calculation obtained in Eq. (\ref{order_2_kac_scaling}).
In both cases, the integrals involved were evaluated numerically using the Mathematica. As expected one observes on this plot that 
the Mean Field calculation is only in qualitative agreement with the numerical results, 
we recall in particular that $\theta_{\rm MF}(2) = 1/2\pi = 0.159155...$. Interestingly, one sees that the second order calculation
is a clear improvement over the Mean Field calculation which is in quite 
good agreement with the numerical results for the scaling function, in particular, $\theta(2) = (\pi +4)/4\pi^2 = 0.180899...$. We have checked 
that this scaling (\ref{fss1}) holds for other values of $d$. In the right panel of Fig. \ref{Fig4}, we show a plot of $n^{\theta(3)}
  P_0([0,x],n)$ for $d=3$ and $\theta(3) = 0.238(4)$ as a function of $n(1-x)$ for different degrees $n=256, 512, 1024$. Again, the value
of $\theta(3) = 0.238(4)$ for which one obtains the best collapse of the curves for different values of $n$ is in good agreement with the values of $\theta(3)$
found for the diffusion equation \cite{persist_diffusion, newman_diffusion}.

\begin{figure}[h]
\includegraphics[angle=-90,width=0.5 \linewidth]{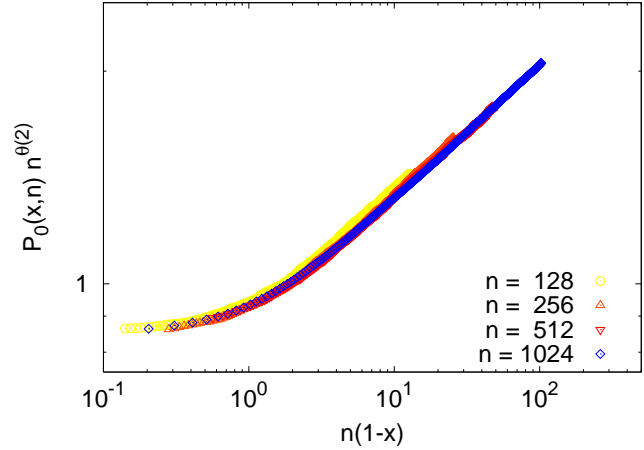}
\caption{Plot of $n^{\theta(2)} \overline{P}_0([0,x],n)$ on a log-log
  scale, where $\overline K_n(x)$ is defined in Eq. (\ref{define_bar}) 
as a function of $n(1-x)$ for $d=3$ and 
for different values of $n=128, 256, 512, 1024$.}\label{Fig5}
\end{figure}

We have also checked numerically our results for the gap probability in the outer intervals (\ref{persist_poly_bis}). For that purpose,
we notice that $P_0([x,\infty],n) = \overline{P}_0([0,1/x],n)$, which is easier to compute numerically, where $\overline{P}_0([a,b],n)$ is 
the gap probability associated to the polynomial $\overline{K}_n(x)$ defined such that $K_n(x) = x^n \overline{K}_n(1/x)$ with
\beq
\overline{K}_n(x) = \sum_{i=0}^{n-1} a_{n-i} (n-i)^{\tfrac{d-2}{4}} x^i + a_0 \,x^n \;.\label{define_bar}
\eeq
Thus, from Eq. (\ref{persist_poly_bis}) one expects that for $0< 1-x \ll 1$, and $n \gg 1$, keeping the product $n(1-x)$ fixed one has 
\beq
\overline{P}_0([0,x],n) \propto n^{-\theta(2)} h^{+}(n(1-x)) \;,\label{persist_bar}
\eeq
independently of $d$. In Fig. \ref{Fig5}, we show a plot of $n^{\theta(2)} \overline{P}_0([0,x],n)$ as a function of $n(1-x)$ for $d=3$ and 
for different values of $n=128, 256, 512, 1024$. Again, the good collapse obtained for $\theta(2)=0.1875$
corroborates the validity of the scaling in Eqs (\ref{persist_poly_bis},  \ref{persist_bar}).

%\begin{figure}[h]
%\begin{minipage}{0.5\linewidth}
%\includegraphics[angle=-90,width=\linewidth]{persist_3d_r.eps}
%\end{minipage}\hfill
%\begin{minipage}{0.5\linewidth}
%\includegraphics[angle=-90,width=\linewidth]{persist_3d_r.eps}
%\end{minipage}
%\end{figure}

\begin{figure}[h]
\begin{minipage}{0.5\linewidth}
\includegraphics[angle=-90,width=\linewidth]{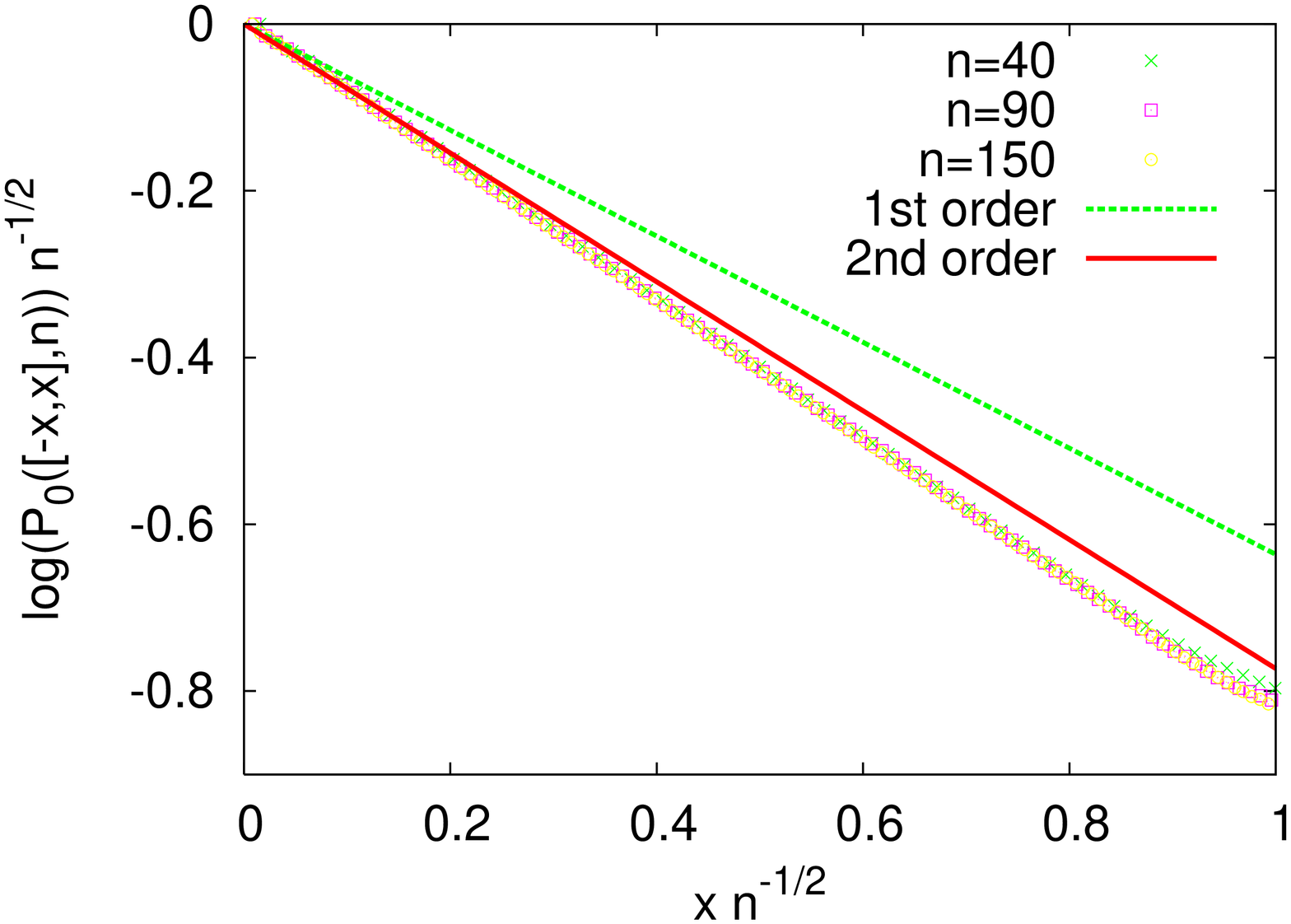}
\end{minipage}\hfill
\begin{minipage}{0.5\linewidth}
\includegraphics[angle=-90,width=\linewidth]{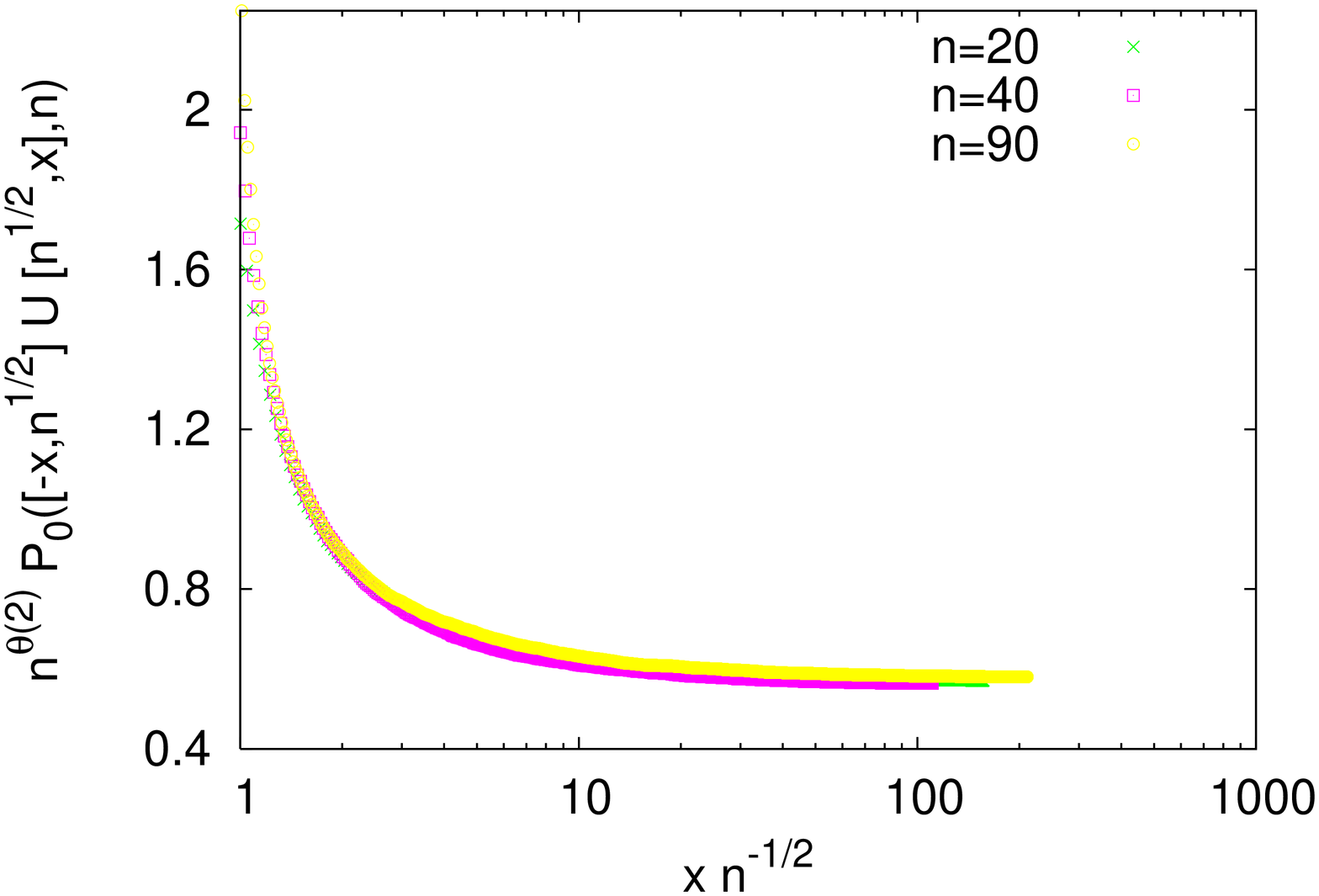}
\end{minipage}
\caption{{\bf Left :} Plot of $[\log P_0([-x,x],n)] n^{-1/2}$ for Weyl
  polynomials $W_n(x)$ (\ref{def_weyl_poly}) as a function of
  $x/\sqrt{n} < 1$ for different values of $n=40, 90$ and $150$. The
  dotted line is the result of the Mean Field approximation whereas
  the solid one is the result of the expansion up to order 
${\cal O}(\epsilon^2)$ in Eq.~(\ref{order_2_weyl}). {\bf Right }: Plot of  $n^{\theta(2) }P_0([-x, -\sqrt{n}] \cup [\sqrt{n},x],n)$
as a function of $x/\sqrt{n} > 1$ for different degrees $n=20,40$ and $90$.}\label{Fig6}
\end{figure}

\begin{figure}[h]
\begin{minipage}{0.5\linewidth}
\includegraphics[angle=-90,width=\linewidth]{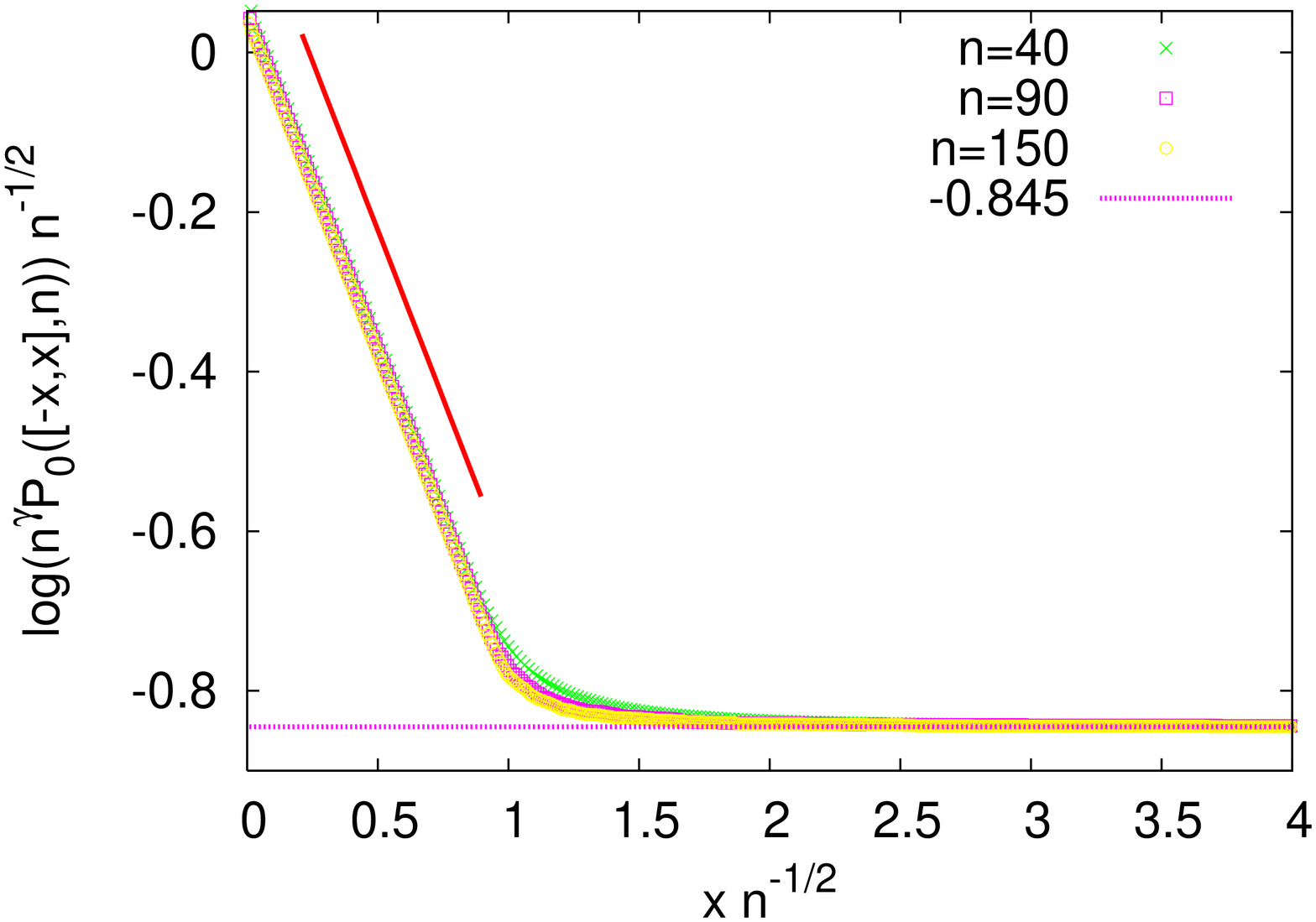}
\end{minipage}\hfill
\begin{minipage}{0.5\linewidth}
\includegraphics[angle=-90,width=\linewidth]{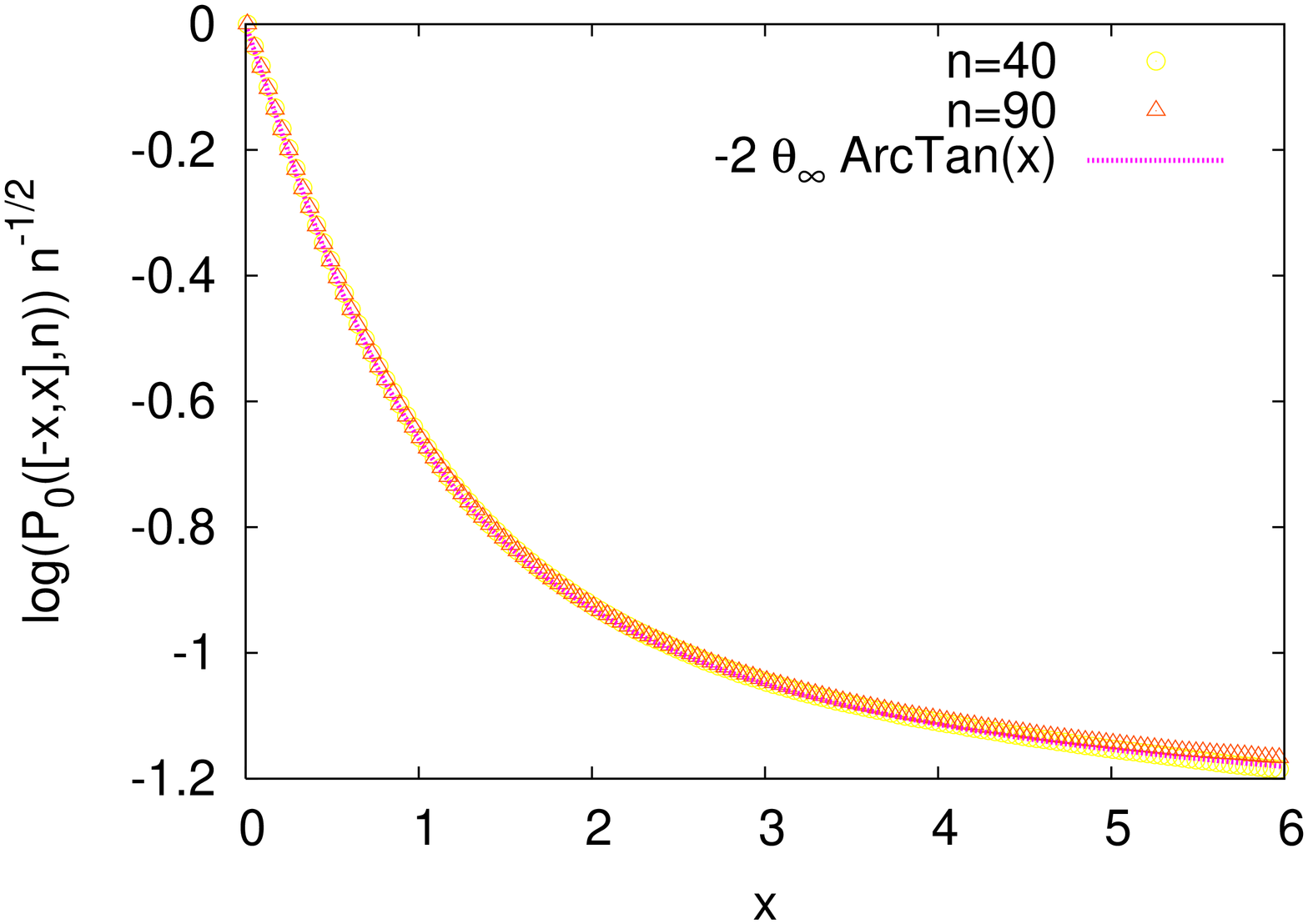}
\end{minipage}
\caption{{\bf Left } : Plot of  $[\log (n^\gamma P_0([-x,x],n))] n^{-1/2}$ for Weyl polynomials $W_n(x)$ (\ref{def_weyl_poly}) as a function of $x/\sqrt{n}$ for different values of $n=40, 90$ and $150$.  
Here $\gamma = 0.10(1)$ and the solid line has slope $-0.845$. {\bf Right } : Plot of  $[\log P_0([-x,x],n)] n^{-1/2}$ for binomial polynomials $B_n(x)$ (\ref{def_binom_poly}) as a function of $x$ for different values of $n=40$ and $90$. The solid line is the exact result in Eq. (\ref{persist_bp_exact}) with $\theta_{\infty} = 0.42(1)$, which is the only fitting parameter here.}\label{Fig7}
\end{figure}

{\bf Weyl polynomials:} We first focus on the inner interval $[-\sqrt{n}, \sqrt{n}]$ and compute numerically the gap probability $P_0([-x,x],n)$ 
for $x < \sqrt{n}$. Here also, this probability was computed by averaging over $10^4$ different realizations of the random coefficients $a_i$'s. 
In the left panel of Fig. \ref{Fig6}, we show a plot of $[\log
  P_0([-x,x],n)] n^{-1/2}$ as a function of $x/\sqrt{n} < 1$ for
different values of $n=40, 90$ and $150$. 
 According to our prediction in Eq. (\ref{persist_weyl_1}), $P_0([-x,x],n)$ behaves exponentially for large $x$. From the slope of the straight line
in the left panel of Fig. \ref{Fig6}, one extracts $2 \theta_{\infty} = 0.845(3)$, in  good agreement with previous numerical estimates from the
persistence probability for the diffusion equation in large dimension \cite{persist_diffusion, newman_diffusion}. On the same figure, we have also
plotted with a dotted line the result from the Mean Field approximation (\ref{weyl_poissonian}) and with a solid line the result up to second order in the $\epsilon$ expansion in Eq. (\ref{order_2_weyl}).
Again, the second order term allows to improve significantly the Mean Field prediction. We recall the estimate up to order~${\cal O}(\epsilon^2)$, 
$2 \theta_{\infty} = 0.7729...$ from Eq. (\ref{proba_second_order_weyl}). We now focus on the outer intervals 
$[-\infty, -\sqrt{n}]$ and $[\sqrt{n},+\infty]$. In the right panel of Fig. \ref{Fig6}, we plot $n^{\theta(2) }P_0([-x, -\sqrt{n}] \cup [\sqrt{n},x],n)$
as a function of $x/\sqrt{n} > 1$ for different degrees $n=20,40$ and $90$. The fact that the curves for different $n$ collapse on a single master curve
is in agreement with the scaling proposed in Eq. (\ref{persist_weyl_3}). 
Finally, we computed the gap probability on the full real axis. In the right panel 
Fig. \ref{Fig7}, we plot $[\log (n^\gamma P_0([-x,x],n))] n^{-1/2}$ as a function of $x/\sqrt{n}$ for different values of $n=40, 90$ and $150$. The exponent
$\gamma = 0.10(1)$, which is the only fitting parameter is fixed to obtained the best collapse of the different curves in the large $x/\sqrt{n}$ limit. 
The solid line has a slope $-2\theta_{\infty} = 0.845$, which is also the value reached by $\log (n^\gamma P_0([-x,x],n)) n^{-1/2}$ for large $x$. This
fact is in complete agreement with the scaling proposed in Eq. (\ref{persist_weyl_entire}). The fact that $\gamma < \theta(2)$ arises from
the correlations between the inner and outer intervals.

{\bf Binomial polynomials:} Finally, we have checked the exact result in Eq.~(\ref{persist_bp_exact}) for binomial polynomials (\ref{def_binom_poly}). 
For that purpose, we have computed numerically $P_0([-x,x],n)$ by averaging over $10^5$ different realizations of the random coefficients $a_i$'s. In the 
right panel of Fig. \ref{Fig7}, we show a plot of $[\log P_0([-x,x],n)] n^{-1/2}$ as a function of $x$ for different values of $n=40$ and $90$. The solid
line is the analytic prediction from Eq. (\ref{persist_bp_exact}) with $\theta_{\infty} = 0.42 (1)$, consistent with our previous estimates from Weyl 
polynomials (see left panel of Fig. \ref{Fig7}). The good agreement with the numerics confirms also the exact result for the probability that these polynomials
have no real root on the real axis in Eq. (\ref{persist_bp_exact_2}).

\subsection{Probability of $k$ real roots : large deviation function}

We now generalize our analysis 
and study the probability $P_k([a,b],n)$ that such polynomials
(\ref{def_kac_poly},~\ref{def_weyl_poly},~\ref{def_binom_poly}) have 
exactly $k$ real roots \cite{dembo} in a given interval $[a,b]$. 

\subsubsection{Mean field approximation : Poisson approximation}

One first considers the Mean Field approximation introduced above where
one assumes that the real 
roots are totally independent and 
randomly distributed with density $\rho_n(x)$. This leads to Eq. (\ref{poissonian_1}) 
\bea\label{poisson}
P_k([a,b],n) = \frac{\langle N_n[a,b]\rangle^k}{k!} e^{-\langle
  N_n[a,b]\rangle} \;.
\eea   
If we focus on the limit $n \gg k \gg 1$, keeping the ratio $y = k/
\langle N_n[a,b]\rangle$ fixed, one has
\beq\label{poisson_scaling}
\log P_k([a,b],n) \sim - \langle N_n[a,b]\rangle \; \varphi_{\rm
    MF}\left(\frac{k}{\langle N_n[a,b]\rangle} \right) \, , \,
\varphi_{\rm MF}(y) = 1 + y\log{y}-y \;,
\eeq
where we have used the Stirling's formula $\log {(k !)} =
(k+\tfrac{1}{2})\log k -k + {\cal O}(1)$. We will see below (through
scaling analysis as well as numerics) that this
Mean Field approximation provides the correct scaling form for
$P_k([a,b],n)$ (although the exact computation of $\varphi(y)$
certainly demands a more sophisticated analysis). Let us present
the consequences of this scaling form in Eq. (\ref{poisson_scaling})
for the different polynomials under study. 

{\bf Kac polynomials:} Let us define $q_{k}(n)=P_k([0,1],n)$. In that case, we have seen in
Eq. (\ref{number_kac_inner}) 
that $\langle N_n([0,1])\rangle~\sim~\tfrac{1}{2 \pi}
\sqrt{\tfrac{d}{2}}  \log n$ so that one expects 
the scaling form
\beq\label{scaling_q_kac}
\log q_k(n) \propto - \log {n} \, \varphi \left(2\pi \sqrt{\frac{2}{d}}\frac{k}{\log n} \right) \;.
\eeq
For the special case of Kac polynomials ($d=2$), this scaling form, 
in the neighborhood of $k=\log n/2\pi$, is consistent with the rigorous
result \cite{maslova} that in this neighborhood $q_k(n)$
is a Gaussian with mean $\log n/2\pi$ and variance 
$V_n~\sim~(\tfrac{1}{\pi}~-~\tfrac{2}{\pi^2})\log n $ (\ref{variance_maslova}) in the
large $n$ limit. 

{\bf Weyl's polynomials:} Let us define $q_k(n) =
P_k([-\infty,\infty],n)$. According to Eq. (\ref{total_weyl}), which
tells us that $\langle N_n [-\infty,\infty] \rangle \sim
(2/\pi)\sqrt{n}$,  
and the scaling form in Eq. (\ref{poisson_scaling}) one expects
\begin{eqnarray}\label{scaling_q_weyl}
\log q_k(n) \propto - \sqrt{n} \, \varphi\left( \frac{\pi}{2}\frac{k}{\sqrt{n}}\right) \;.
\end{eqnarray}

{\bf Binomial polynomials:} Let us define similarly $q_k(n) =
P_k([-\infty,+\infty],n)$. In that case, according to
Eq. (\ref{total_bp})  which
tells us that $\langle N_n [-\infty,\infty] \rangle = \sqrt{n}$,   
and the scaling form in Eq. (\ref{poisson_scaling}) one expects
\begin{eqnarray}\label{scaling_q_bp}
\log q_k(n) \propto - \sqrt{n} \, \varphi\left( \frac{k}{\sqrt{n}}\right) \;.
\end{eqnarray}
In the following, we will check these scaling forms (\ref{scaling_q_kac}-\ref{scaling_q_bp}) numerically. 

\subsubsection{A more rigorous approach for a smooth Gaussian stationary process}

We illustrate this approach on the diffusion equation with random initial conditions (\ref{diff_eq}), which is the underlying
stochastic process describing the statistics of real roots of these random polynomials. We thus consider
the probability $p_k(t,L)$ that the diffusing field $\phi({\mathbf x},t)$ crosses zero exactly
$k$ times up to time $t$. Let us first consider the regime
$1~\ll~t~\ll~L^2$. In this regime, $p_k(t,L)$
is given by the probability ${\cal P}_k(T)$ that $X(T)$ crosses zero
exactly $k$ 
times where $X(T)$ is a GSP with correlations $a(|T-T'|) =
[{\rm cosh}(|T-T'|/2)]^{-d/2}$, where $T=\log t$. 
Since, $a(T) = 1 - \tfrac{d}{16} T^2 + o(T^2)$ for small $T$,
this GSP is a smooth process with a finite density of zero crossings 
given by the
Rice's formula $\mu = [-{a''(0)}]^{{1}/{2}}/\pi$
\cite{rice_formula}. We propose the following scaling form for
large $T$ and large~$k$, with $k/T$ fixed
\begin{eqnarray}
\log{{\cal P}_k(T)} = -T \varphi\left(\frac{k}{\mu T}
\right). 
\label{scaling}
\end{eqnarray} 
To understand the origin of this scaling form, let us
consider the generating function $\hat {\cal P}(p,T) = \sum_{k=0}^\infty p^k
{\cal P}(k,T)$ as in Eq. (\ref{generating_function}). One can show \cite{satya_partial} that $\hat {\cal P}(p,T) \sim \exp[-\hat \theta(p) T]$,
where for a smooth GSP $\hat \theta(p)$ depends
continuously on $p$. If the scaling in Eq. (\ref{scaling}) holds, one gets by steepest
descent method valid for large $T$, $\hat \theta(p) =
{\rm Min}_{x > 0} 
[\mu x \log p - \varphi(x)]$. Inverting the Legendre transform we get 
\begin{eqnarray}
\varphi(x) = {\rm Max}_{0 \leq p \leq 2}[ \mu x \log p + \hat \theta(p)] \;.
\label{legendre} 
\end{eqnarray}
Notice that although $\hat \theta(p)$ is a priori defined on the interval
$[0,1]$, the computation of $\varphi(x)$ involves an
analytical continuation of $\hat \theta(p)$ on $[0,2]$. Going back to real
time $t$, Eq. (\ref{scaling}) then yields a rather unusual scaling
form valid in the 
limit $1\ll t \ll L^2$ 
\begin{eqnarray}
\log p_k(t,L) \sim -\log{t} \, \varphi\left(\tfrac{k}{\mu \log t} \right) \;.
\label{peculiar_scaling} 
\end{eqnarray}
In the opposite limit $t\gg L^2$, one simply replaces $t$ in
(\ref{peculiar_scaling}) by $L^2$. 
Translating into random polynomials, this regime corresponds to
 $(1-x) \ll n^{-1}$ since one just replaces $t$ by $1/(1-x)$ and 
$L^2$ by the degree $n$ as discussed before. Hence, 
in this regime, we arrive 
at the announced scaling
form for $q_k(n)$ in Eq. (\ref{scaling_q_kac}). This approach can be
extended straightforwardly to the other classes of 
polynomials, yielding the scaling forms in Eq. (\ref{scaling_q_weyl},
\ref{scaling_q_bp}).  

Of course, despite the exact formula (\ref{legendre}), the function
$\varphi(x)$ remains very hard to compute, simply because $\hat
\theta(p)$ is, in many cases, unknown. However, for the random
acceleration process (RAP), sometimes called in the literature
``integrated brownian motion'' 
\beq\label{def_RAP}
\frac{d^2 \; x(t)}{d\,t^2} = \eta(t) \;, \; \langle \eta(t)
\eta(t')\rangle = \delta(t-t') \;,
\eeq
where $\eta(t)$ is a white noise for
which $\mu_{\rm RAP} = \sqrt{3}/(2\pi)$, $\hat \theta_{\rm RAP}(p)$ has been computed exactly
\cite{burkhardt, desmedt}, yielding $\hat \theta_{\rm RAP}(p) =
\tfrac{1}{4}[1-\tfrac{6}{\pi}\sin^{-1}(\tfrac{p}{2})]$. By performing the
Legendre transform (\ref{legendre}) one obtains 
\bea
\varphi_{\rm RAP}(x) = \frac{\sqrt{3}}{2\pi} x
\log{\left(\frac{2x}{\sqrt{x^2+3}} 
  \right)} + \frac{1}{4} \left(1-\frac{6}{\pi} {\rm
    ArcSin}\left(\frac{x}{\sqrt{x^2+3}}\right) \right) \;,
\eea
with the asymptotic behaviors 
\begin{eqnarray}\label{large_dev_rnd_acc}     
\varphi_{\rm RAP}(x) \sim
\begin{cases}
\tfrac{1}{4} + \tfrac{\sqrt{3}}{2\pi} x \log x \quad, \quad x \to 0 \\
\tfrac{3\sqrt{3}}{16 \pi} (x-1)^2 \quad, \quad x \to 1 \\
\tfrac{\sqrt{3}}{2\pi} x \log{2} \quad, \quad x \to \infty \;,
\end{cases}
\end{eqnarray}
which gives back the exact result $\varphi_{\rm RAP}(0) = 1/4$
\cite{sinai, burkhardt_theta}.

\subsubsection{Numerical results}

\begin{figure}[h]
\begin{minipage}{0.5\linewidth}
\includegraphics[angle=-90,width=\linewidth]{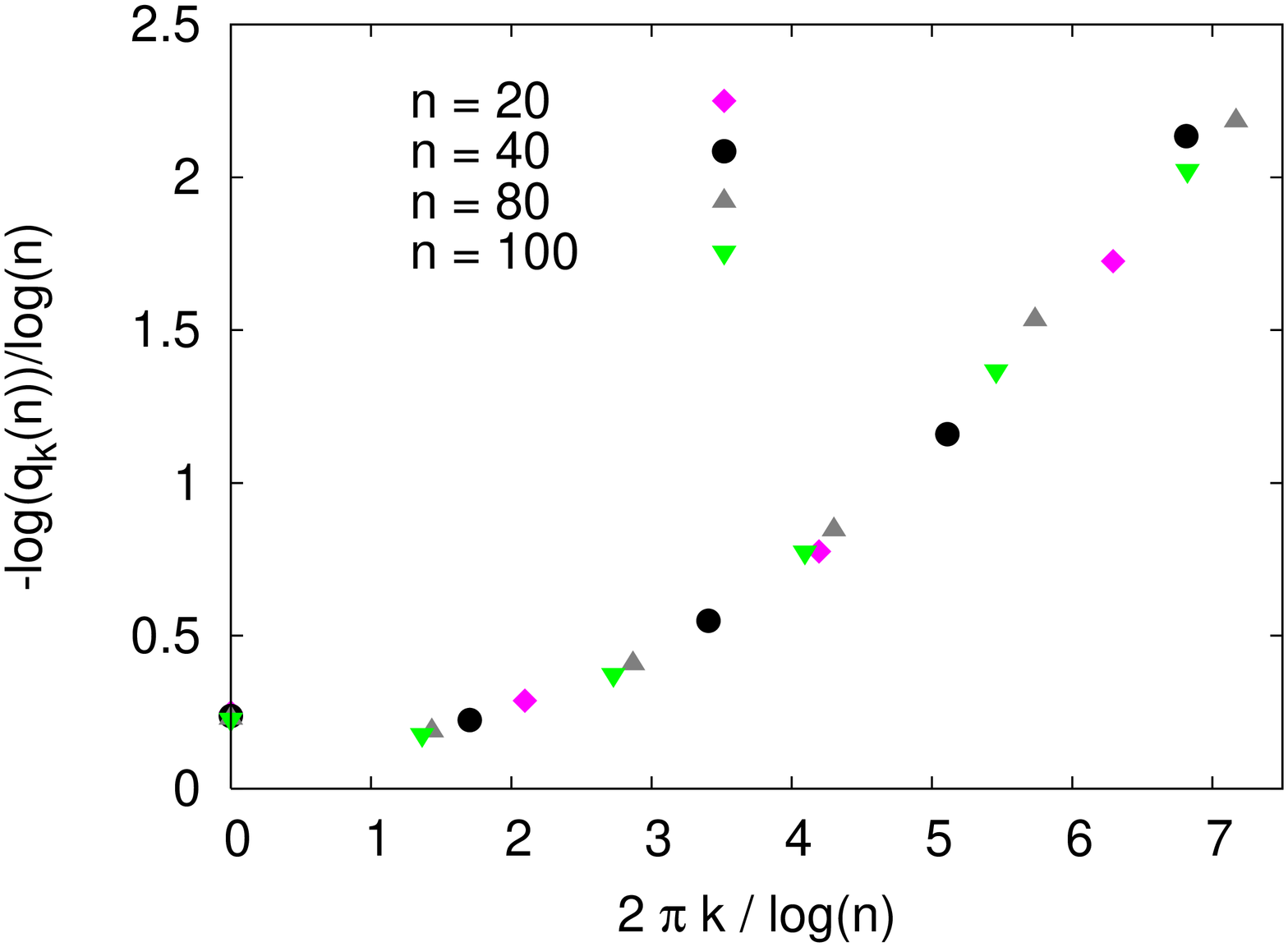}
\end{minipage}\hfill
\begin{minipage}{0.5\linewidth}
\includegraphics[angle=-90,width=\linewidth]{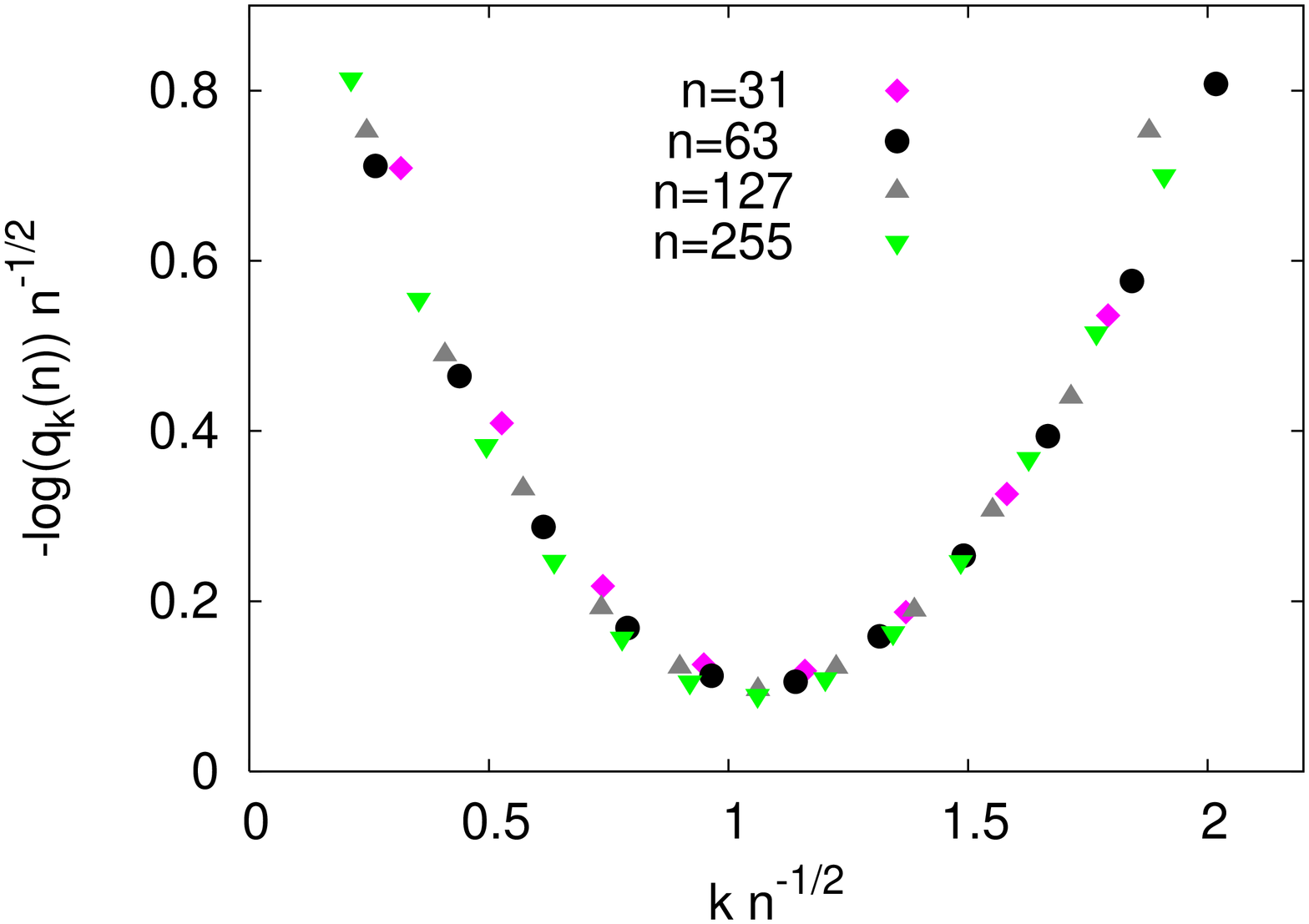}
\end{minipage}
\caption{{\bf Left} : Plot of $-\log(q_k(n)/\log{n})$ as a function
of $2\pi k/\log{n}$ for Kac polynomials (\ref{def_kac_poly}) with $d=2$ and for different values of $n=20, 40, 80$ and $100$. {\bf Right} : Plot of $-\log{(q_k(n))} n^{-1/2}$ as a function of $k/\sqrt{n}$ for binomial polynomials (\ref{def_binom_poly}) for different values of $n=31, 63, 127$ and $255$.
}\label{Fig8}
\end{figure}

In Ref.~\cite{us_prl}, we have checked numerically the scaling form (\ref{peculiar_scaling}) for the diffusion equation with random initial conditions. Here, 
we have computed numerically these probabilities $q_k(n)$ for the different polynomials under study. This was done by averaging over
$10^4$ different realizations of the random coefficients $a_i$'s. In the left panel of Fig.~\ref{Fig8}, we show a plot of $-[\log{q_k(n)}]/\log{n}$ as a function
of $2\pi k/\log{n}$ for Kac polynomials (\ref{def_kac_poly}) with
$d=2$ and for different values of $n=20, 40, 80$ and $100$. This
suggests that the different points fall on a single master curve,
which is in rather good agreement with the scaling form proposed in Eq. (\ref{scaling_q_kac}). Similarly, in the right panel
of Fig. \ref{Fig8}, we show a plot of $-[\log{q_k(n)}] n^{-1/2}$ as a
function of $k/\sqrt{n}$ for binomial polynomials
(\ref{def_binom_poly}) for different values of $n=31, 63, 127$ and
$255$. Here also, the fact the points fall a single master curve is in
good agreement with the scaling form in Eq. (\ref{scaling_q_bp}). We have also checked that a similar scaling, as in Eq. (\ref{scaling_q_weyl}) holds for Weyl polynomials (\ref{def_weyl_poly}).

\subsection{Distribution of the maximum of real roots}

Up to now, we have mainly focused on the distribution of the minimum
$\lambda_{\rm min}$ 
of the absolute values $\{|\lambda_i|\}$ of the real roots of these
polynomials. Indeed, the gap probability  
$P_0([-x,x],n)$  
is just the probability that $\lambda_{\rm min}$ is larger than $x > 0$. We now instead focus
on the distribution of the {\it maximum} $\lambda_{\rm max}$ of the $\{ |\lambda_{i}|\}$. 

\subsubsection{Mean Field approximation}

As a first approach, we consider the Mean Field or Poissonian approximation introduced above, where one neglects the correlations between the real roots. Then, for any random polynomial ${Q}_n(x) = \sum_{i=0}^n b_i x^i$, the probability that $\lambda_{\rm max} \leq x$ is simply 
the probability that ${Q}_n(x)$ has no real root in $[-\infty, -x] \cup [x,+ \infty]$. Therefore, within the Mean Field approximation one has
\beq\label{max_mf}
{\rm Proba.}\;(\lambda_{\rm max} \leq x,n )= \exp{\left(-2\int_x^\infty \rho_n(t) dt  \right)} \;.
\eeq
Taking the derivative of this expression (\ref{max_mf}) with respect to $x$, one obtains the probability distribution function~(pdf) $p_{{\rm max}}(x,n)$ of
the maximum $\lambda_{\rm max}$ as
\beq\label{max_pdf_mf}
p_{{\rm max}}(x,n) = 2 \rho_n(x) \exp{\left(-2\int_x^\infty \rho_n(t) dt  \right)} \;.
\eeq
To obtain the large $x$ behavior of $p_{{\rm max}}(x,n) $ one computes the one of $\rho_n(x)$ whose expression is given by
\beq\label{expr_rho_max}
 \rho_n(x) = \frac{\sqrt{c_n(x) (c_n'(x)/x + c_n''(x)) - [c_n'(x)]^2   }}{2 \pi c_n(x)} \;, \; c_n(x) = \sum_{i=0}^n \langle b_i^2\rangle \, x^{2i} \;.
\eeq
For large $x$, one has $c_n(x) = x^{2n} \langle b_{n}^2\rangle + x^{2n-2}  \langle b_{n-1}^2\rangle + {\cal O}(x^{2n-4})$ which gives $\rho_n(x) \sim \sqrt{{\langle b_{n-1}^2 \rangle }/{\langle b_n^2 \rangle}}/(\pi x^2)$. Finally, from Eq. (\ref{max_pdf_mf}), one obtains for $x \gg 1$
\begin{eqnarray}\label{max_tail_mf}
p_{{\rm max}}(x,n)  \sim 2 \rho_n(x) \sim \frac{2}{\pi x^2}\sqrt{\frac{\langle b_{n-1}^2 \rangle }{\langle b_n^2 \rangle}} \;.
\end{eqnarray}

\subsubsection{Exact result for the tail}

In fact, the tail of the distribution can be computed exactly by noting that
\beq\label{max_exact}
{\rm Proba.}(\lambda_{\rm max} \leq x,n )= \overline{P}_0\left(\left [-\frac{1}{x},\frac{1}{x}\right ],n\right) \, ,
\eeq
where $\overline{P}_0([a,b],n)$ is the gap probability associated to the polynomial 
$\overline{Q}_n(x)$ defined such that ${Q}_n(x) = x^n \overline{Q}_n(1/x)$. Similarly, we denote $\overline{\rho}_n(x)$ the
mean density of real root associated to $\overline{Q}_n(x)$. 
From this exact identity (\ref{max_exact}), valid for all polynomials ${Q}_n(x)$ and all $n \geq 1$, 
one obtains the asymptotic behavior
\beq\label{max_asympt}
{\rm Proba.}(\lambda_{\rm max} \leq x,n )  = 1 - 2 \frac{\overline{\rho}_n(0)}{x} + {o(x^{-1})} \, ,
\eeq
where we have simply used the definition of $\overline{\rho}_n(0)$, 
provided it is well defined, which is the case for Gaussian random polynomials. From this asymptotic behavior (\ref{max_asympt}), one obtains the 
pdf $p_{{\rm max}}(x,n)$ of
the maximum $\lambda_{\rm max}$ for large $x > 0$ as 
\beq\label{max_asympt_pdf}
p_{{\rm max}}(x,n) \sim 2 \frac{\overline{\rho}_n(0)}{x^2} \; , \;  \overline{\rho}_n(0) =
\frac{1}{\pi} \sqrt{\frac{\langle b_{n-1}^2 \rangle }{\langle b_n^2 \rangle}}
\eeq
where we have used the formula (\ref{expr_rho_max}) to compute $\overline{\rho}_n(0)$. For the various polynomials under consideration here, one thus obtains such a power law tail (\ref{max_asympt_pdf})
with $\overline{\rho}_n(0) = \pi^{-1} ((n-1)/n)^{\tfrac{d-2}{4}}$ for Kac polynomials, $\overline{\rho}_n(0) = 
\sqrt{n}/\pi$ for Weyl polynomials and binomial polynomials. This result shows in particular that the mean value of $\lambda_{\rm max}$ is
not defined for these polynomials. 

It is interesting to note that the
Mean Field approximation (\ref{max_tail_mf}) gives the exact result
for this algebraic tail (\ref{max_asympt_pdf}). This might be
understood by noting that, apart from a short range
repulsion, the real roots of these polynomials are actually weakly
correlated. For instance, for Weyl polynomials, we show in Appendix \ref{appendix_scaling_weyl}, see Eq. (\ref{asympt_tilde}), that 
the two-point (connected) correlation function of the real roots inside the
interval $[-\sqrt{n}, \sqrt{n}]$ decays faster than exponentially at
large distance (see also Ref.~\cite{bleher} for similar properties of
$K_n(x)$ and $B_n(x)$). On the other hand, the marginal distribution
of a single real root is nothing else but $\rho_n(x)/\langle
N_n[-\infty, \infty]\rangle$ which decays algebraically $\rho_n(x) \sim 1/x^2$ for large $x$. Given that these real roots are weakly correlated, 
we expect that the distribution of the maximum of these real roots is
given by a Fr\'echet distribution, which indeed has a power law tail,
as we found here (\ref{max_asympt_pdf}).

\subsubsection{Numerical results}

\begin{figure}[h]
\begin{minipage}{0.5\linewidth}
\includegraphics[angle=-90,width=\linewidth]{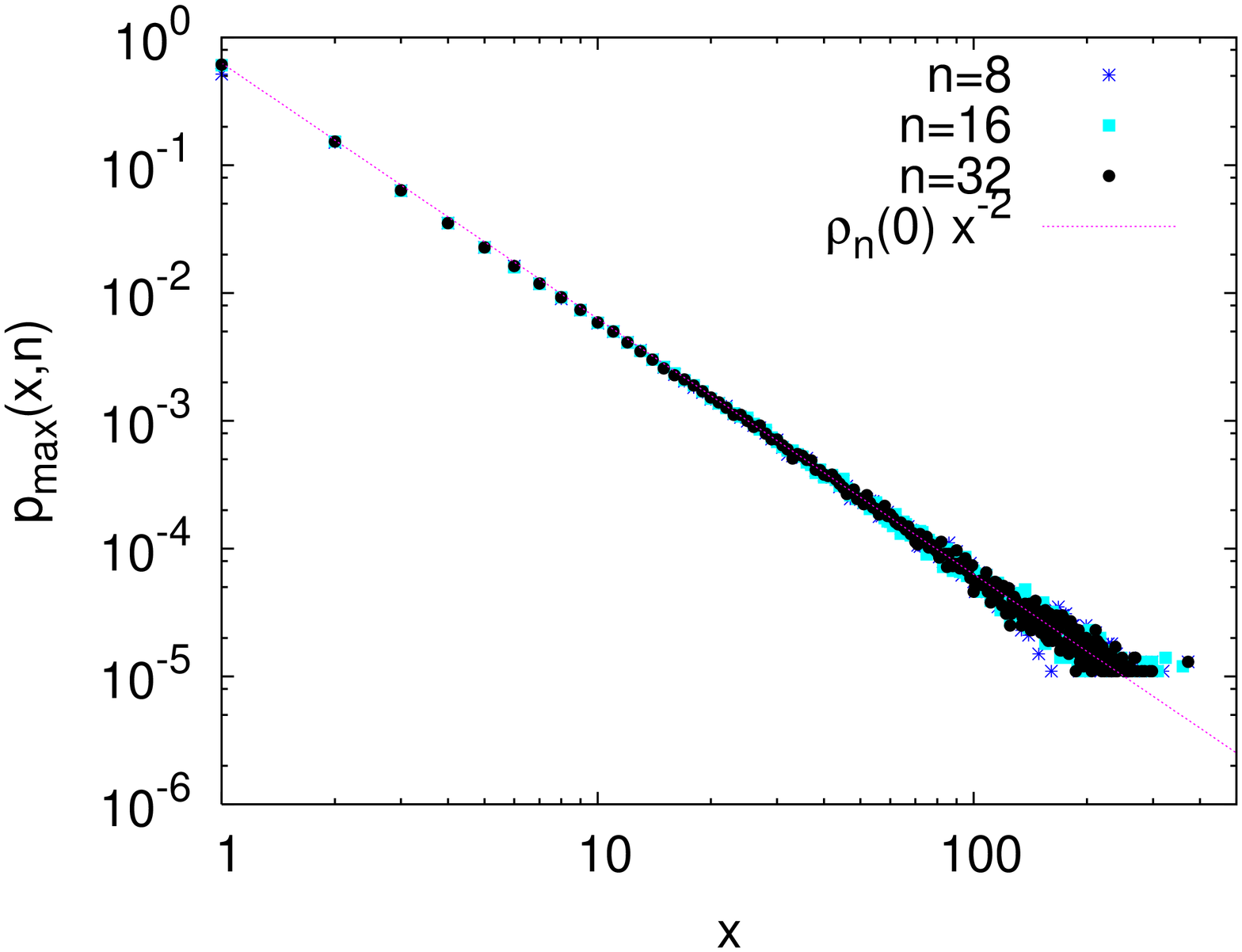}
\end{minipage}\hfill
\begin{minipage}{0.5\linewidth}
\includegraphics[angle=-90,width=\linewidth]{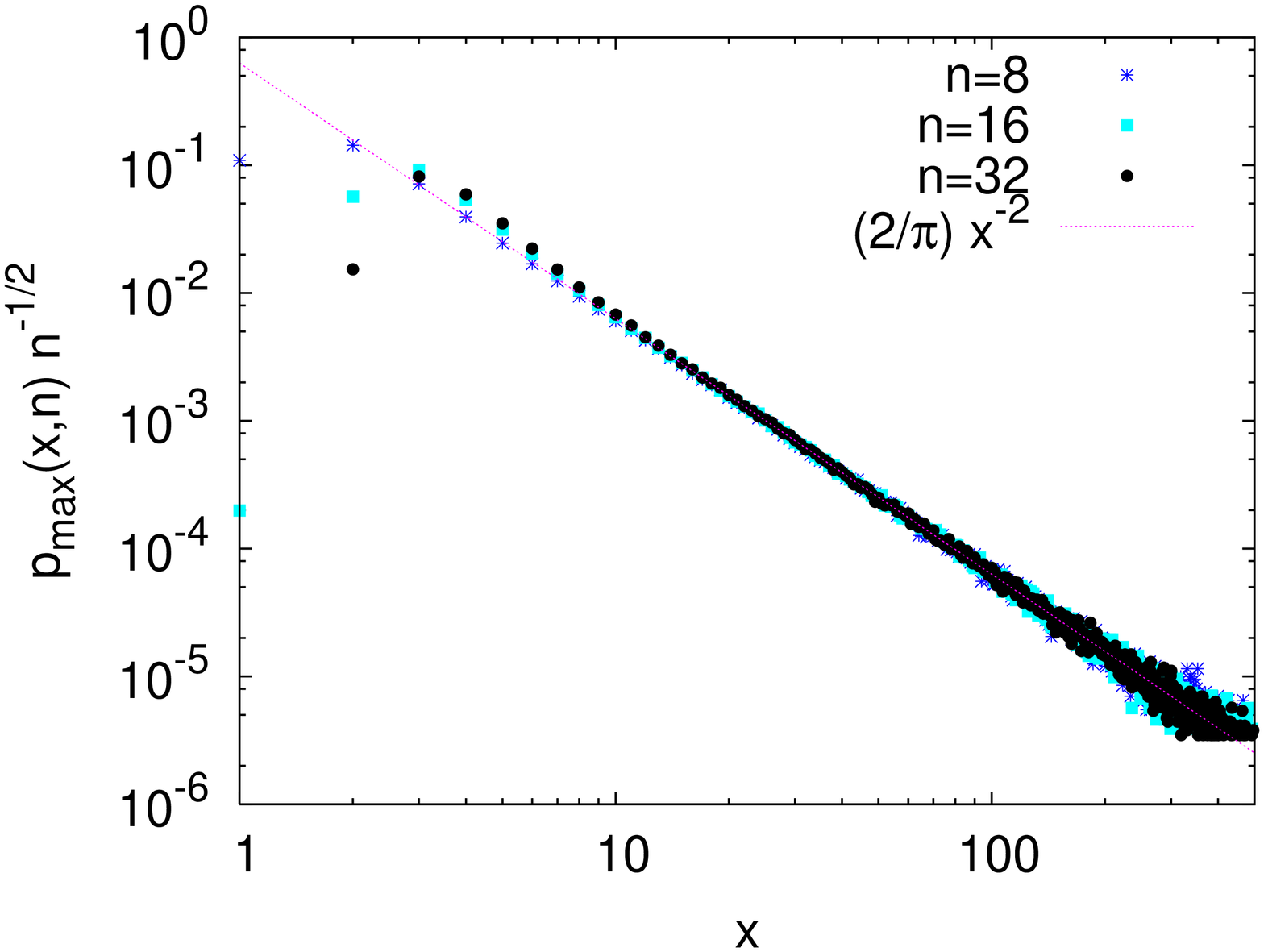}
\end{minipage}
\caption{{\bf Left :} $p_{{\rm max}}(x,n)$ for Kac polynomials
  $K_n(x)$ (\ref{def_kac_poly}) and $d=2$ as a function of $x$ for different
  values of $n=8, 16$ and $32$. The dotted line is the exact result
  for the tail of the distribution (\ref{max_asympt_pdf}). {\bf Right
  } : $p_{{\rm max}}(x,n)/\sqrt{n}$ for Weyl polynomials
  $W_n(x)$ (\ref{def_weyl_poly}) as a function of $x$ for different
  values of $n=8, 16$ and $32$. The dotted line is the exact result
  for the tail of the distribution (\ref{max_asympt_pdf}) }\label{Fig9}
\end{figure}

We have checked numerically this exact asymptotic behavior
(\ref{max_asympt_pdf}) for the three classes of random polynomials~(\ref{def_kac_poly}, \ref{def_weyl_poly}, \ref{def_binom_poly}). In all the three cases the
pdf of the maximum $\lambda_{\rm max}$ was obtained by averaging
over $10^5$ different realizations of the random Gaussian variables $a_i$'s. In the left panel of Fig.~\ref{Fig9}, we
show a plot of $p_{{\rm max}}(x,n)$ as a function of $x$ for Kac
polynomials and $d=2$, for different values of $n=8, 16$ and
$32$. Notice that the exact result 
in Eq.~(\ref{max_asympt_pdf}), which is plotted here with a dotted line, is in principle true for all $n \geq 1$ so that it is not necessary to perform numerics for polynomials of large degree. This figure shows a good agreement between our analytic result and the numerics. Similarly, in the right panel of
Fig.~\ref{Fig9}, we have plotted $p_{{\rm max}}(x,n)/\sqrt{n}$ for
Weyl polynomials $W_n(x)$ and for different
  values of $n=8, 16$ and $32$. Here again, the agreement with the analytical result in Eq. (\ref{max_asympt_pdf}) is quite good. 

\section{Conclusions and perspectives}

\begin{figure}[h]
\includegraphics[angle=0,width= 0.7 \linewidth]{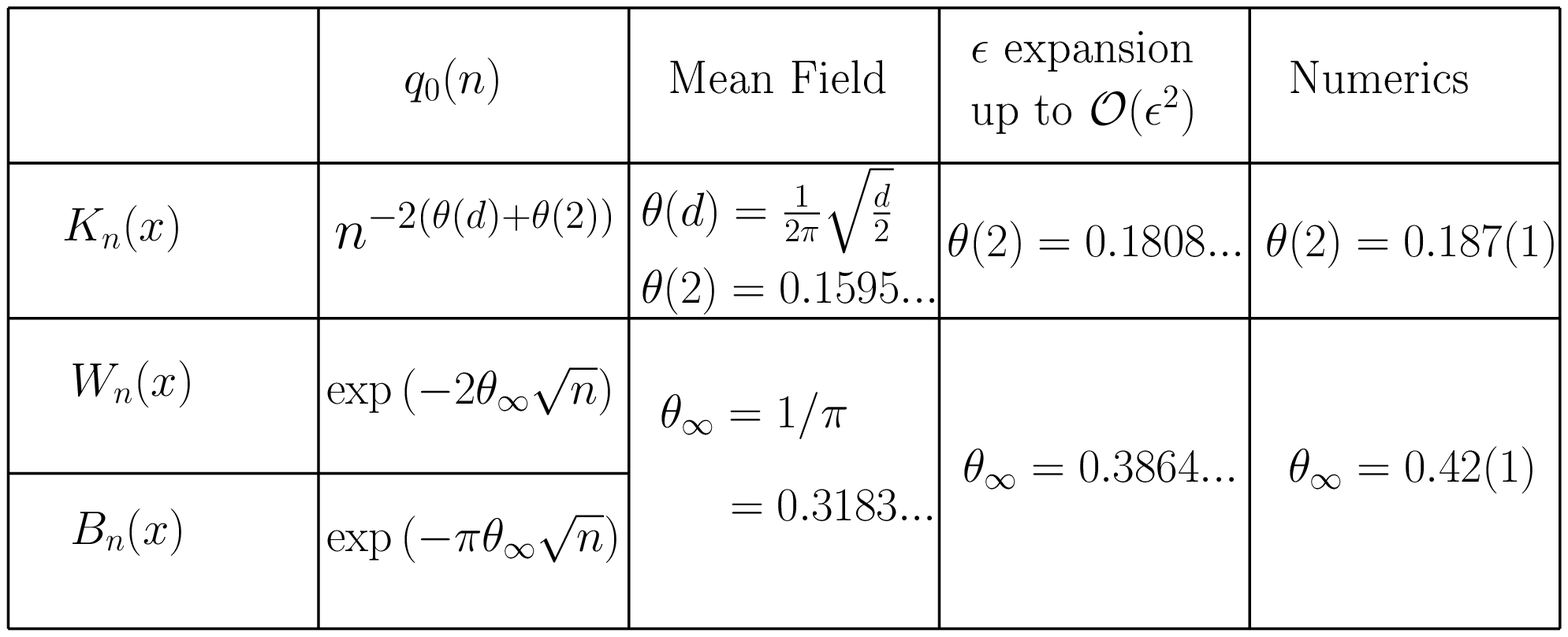}
\caption{Summary of the main results for the probability of no root on the full real axis $q_0(n)$ for the three different classes of polynomials $K_n(x), W_n(x)$ and $B_n(x)$.}\label{fig10}
\end{figure}

In conclusion, we have investigated different statistical properties of the real roots of three different types
of real random polynomials (\ref{def_kac_poly}, \ref{def_weyl_poly}, \ref{def_binom_poly}). In these three classes,
the mean density of real roots exhibit a rich variety of behaviors, as shown in Figs \ref{Fig1}-\ref{Fig3}. We have first 
focused on gap probabilities (\ref{exact_no_root_kac}, \ref{persist_weyl_entire}, \ref{persist_bp_exact_2}) which were shown to be closely 
related to the persistence probability for the diffusion equation with random initial conditions in dimension $d$ (\ref{diff_eq}, \ref{fss1}).  We proposed
a Mean Field approximation to compute the exponents as well as the universal scaling functions describing these gap probabilities. Furthermore,
we showed how to improve systematically this MF approximation (see Fig. \ref{Fig4}, \ref{Fig6}) using an $\epsilon$-expansion based 
on the so called persistence with partial survival. In the case of binomial polynomials $B_n(x)$ (\ref{def_binom_poly}), this allows to obtain exact results for the gap probability (\ref{persist_bp_exact}). Our main results for the gap probability $q_0(n)$ on the full real axis are summarized in Fig. \ref{fig10}. 
We hope
that these connections may help to obtain exact results for the exponent $\theta(d)$, which remains a challenge.
Besides, we extended our analysis to the probability that these random polynomials have exactly $k$ real roots in a given interval $[a,b]$. We have
shown, in the three cases,  that this probability has an interesting scaling form characterized by a large deviation function (\ref{scaling_q_kac}-\ref{scaling_q_bp}). We proposed a Mean Field approximation which reproduces qualitatively these scaling forms, which were further checked numerically (see Fig. \ref{Fig8}). A similar question was asked in the past for 
Ginibre real matrices : what is the probability that such $n \times n$ matrices has exactly $k$ real roots~? Quite recently, 
in Ref.~\cite{kanzieper_akemann}, Akemann and Kanzieper obtained an exact formula for this probability. In that case, the mean number of
real roots grows like $\sqrt{n}$ and our MF approximation would therefore suggest a scaling form for this probability similar to Eq. (\ref{scaling_q_weyl}). It would be very interesting to extract the large deviation function from an asymptotic analysis of the exact result of Ref.~\cite{kanzieper_akemann}. 

Finally, we computed the pdf of the largest real root of these random polynomials. We showed that for a wide class of random polynomials, 
this pdf has an algebraic tail with exponent $-2$ as shown in Eq. (\ref{max_asympt_pdf}) and it would certainly be interesting to extend these results to the case of non-Gaussian random polynomials. 

\begin{acknowledgements}
We thank P. Krapivsky for stimulating discussions. One of us (G.S.) would like to thank the physics
department of the Saarland University for hospitality and generous allocation of computer time.
\end{acknowledgements}

\newpage

\begin{appendix}

\section{Calculation of the mean density for generalized Kac's
  polynomials}\label{appendix_density_kac}  

In this appendix, we give some details concerning the computation of
the mean density and the mean number of real roots for Kac's
polynomials $K_n(x)$ (\ref{def_kac_poly}). 

\subsection{Scaling form} 

The starting point of the calculation is the formula for the density
$\rho_n(x)$ given in Eq. (\ref{ek_formula}), which using
Eq. (\ref{correl_kac}) gives 
\begin{eqnarray}
&&\rho_n(x) =  \frac{1}{\pi} \sqrt{ \frac{s_{d/2+1,n}(x)}{1+x^2
    s_{d/2-1,n}(x)} - \left(\frac{s_{d/2,n}(x)}{1+s_{d/2-1}(x)} \right)^2
    } \;,\label{density_kac_app} \\ 
&& \text{with} \quad s_{k,n}(x) = \sum_{i=1}^n i^k x^{2i-2} \;. \nonumber
\end{eqnarray}
In the limit $n \gg 1$, and $|1-x| \ll 1$ keeping $y = n(1-x)$ fixed,
$s_{k,n}(x)$ can be viewed as a Riemann sum, thus
\begin{eqnarray}
s_{k,n}(x) \sim n^{k+1} \int_0^1 dx x^k e^{-2 u x} \quad, \quad u =
n(1-x) \;.\label{asympt_formula_app}
\end{eqnarray}
Finally using Eq. (\ref{asympt_formula_app}) together with
Eq. (\ref{density_kac_app}) yield the formula
(\ref{scaling_density_kac}) given in the text :
\begin{eqnarray}
&&\rho_n(x) = n \rho^K(n(1-x)) \;, \; \rho^K(y) = \frac{1}{\pi} \sqrt{ \frac{I_{d/2+1}(y)}{I_{d/2-1}(y)} -
  \left(\frac{I_{d/2}(y)}{I_{d/2-1}(y)} \right)^2 } \;, \\
&&I_m(y) = \int_{0}^{1} dx\; x^m \exp{(-2 y x)} \;. \label{def_im_app}
\end{eqnarray}

\subsection{Asymptotic expansions}

To compute the asymptotic behaviors of the function $\rho^K(y)$
in Eq. (\ref{scaling_density_kac}), one needs the asymptotic behaviors
of $I_m(y)$ defined above (\ref{def_im_app}).

\subsubsection{The limit $y \to\infty$}

In that limit, one writes $I_m(y)$ as
\begin{eqnarray}
I_m(y) &=& y^{-1-m} \int_0^y du u^m e^{-2u} = y^{-1-m} \int_0^\infty
du u^m e^{-2u} + {\cal O}(y^m e^{-2y}) \nonumber \\ 
&=& (2 y)^{-1-m} \Gamma(1+m) + {\cal O}(y^m e^{-2y}) \;, \label{ik_large_uplus}
\end{eqnarray}
which, together with Eq. (\ref{scaling_density_kac}), gives the first
line of the formula given in Eq. (\ref{density_large_uplus}) in the text.   

\subsubsection{The limit $y \to -\infty$} 

Because of several cancellations occurring in $\rho^K(y)$ in the
limit $y \to -\infty$ one needs the three first terms in the
asymptotic behavior of $I_m(y)$ in that limit :
\begin{eqnarray}
I_m(y) &=& |u|^{-1-m} \int_0^{|y|} du \; u^m e^{2u} \nonumber \\
&=& e^{2 |y|}\left(\frac{1}{2 |y|} - \frac{m}{4 y^2} + \frac{m(m-1)}{8
  |y|^3}  + {\cal O}(y^{-4})\right) \;,\label{ik_large_uminus}
\end{eqnarray}
which after a tedious but straightforward calculations, using
Eq. (\ref{scaling_density_kac}), gives the second line of the formula
given in Eq. (\ref{density_large_uminus}) in the text.

\section{Calculation of the mean density for real Weyl's
  polynomials}\label{appendix_density_weyl}

In this section, we present the details of the calculation leading to
the scaling form given in Eq. (\ref{density_scaling_weyl}). We start with the expression for the mean density given in Eq. (\ref{density_weyl}) with
$x = u \sqrt{n}$:
\begin{eqnarray}
\rho_n(x) = \rho_n(\sqrt{n} u) = \frac{1}{\pi}\sqrt{1 + \frac{n^{n+1}
    u^{2n}(u^2-1-n^{-1})}{e^{nu^2} \Gamma(n+1,nu^2)} -
    \frac{n^{2n+1}u^{4n+2}}{(e^{nu^2} \Gamma(n+1,nu^2))^2}} \;.
    \label{density_weyl_app1}  
\end{eqnarray}
We want to obtain the limit $n \to \infty$, keeping $u$ fixed, in the
above equation (\ref{density_weyl_app1}) . For that purpose, we need
the large $n$ behavior of $\Gamma(n,n u^2)$ for large $n$ and $u$
fixed. One writes
\begin{eqnarray}
\Gamma(n,nu^2) = n^n \int_{u^2}^\infty \frac{dx}{x} \; e^{-n(x-\log{x})}
= n^n \int_{\log{u^2}}^\infty dy \; e^{-n f(y)} \quad, \quad f(y) = y -\log{y} \;, \label{density_weyl_app2}
\end{eqnarray}
and under this form (\ref{density_weyl_app2}) we can now
perform a saddle-point calculation. The function $f(y)$ has a single
minimum at $y = 1$ and therefore we expect that the large $n$
behavior of the expression in Eq. (\ref{density_weyl_app2}) will
depend on the sign of 
$u-1$. For $u < 1$ the minimum of $f(y)$ lies in the interval of
integration in Eq. (\ref{density_weyl_app2}) and one gets a result independent of $u$
\begin{eqnarray}
\Gamma(n+1,nu^2) = \sqrt{2 \pi }n^{n+1/2} e^{-n}\left(1 + {\cal
  O}(n^{-1/2}) \right) \quad, \quad u < 1 \;. \label{expansion_gamma_smallu}
\end{eqnarray}
On the other hand, for $u > 1$ the minimum of $f(y)$ in
Eq. (\ref{density_weyl_app2}) lies outside the interval of integration
and one gets instead
\begin{eqnarray}
\Gamma(n+1,nu^2) = n^{n-1} e^{-nu^2} \frac{u^{2n}}{u^2-1}
\left(1-\frac{1}{n} \frac{u^2}{(u^2-1)^2} + {\cal O}(n^{-2}) \right)
\quad, \quad u > 1 \;. \label{expansion_gamma_largeu} 
\end{eqnarray}
Using the asymptotic expansion for $u <1$
(\ref{expansion_gamma_smallu}) together with
Eq. (\ref{density_weyl_app1}) one obtains straightforwardly that for
$u < 1$ fixed and large $n$
\begin{eqnarray}
\rho_n(\sqrt{n} u) = \frac{1}{\pi} + {\cal O}(n^{-1/2}) \quad, \quad u
< 1 \;,
\end{eqnarray}
yielding, in the limit $n \to \infty$ the expression
(\ref{density_scaling_weyl}) given in the 
text. Similarly, using Eq. (\ref{expansion_gamma_largeu}) together
with Eq. (\ref{density_weyl_app1}), one gets for $u>1$ fixed and large
$n$
\begin{eqnarray}
\rho_n(\sqrt{n} u) = {\cal O}(n^{-1}) \quad, \quad u > 1 \;,
\end{eqnarray}
yielding, in the limit $n \to \infty$ the expression
(\ref{density_scaling_weyl}) given in the 
text.

\section{Computation of $\langle N^2_n([0,x])\rangle_c$ for Kac's
  polynomials and $d=2$}\label{appendix_scaling_kac} 

\subsection{Scaling form}

In this appendix, we give the details of the computation of $\langle N^2_n([0,x])\rangle$ which lead to 
formula (\ref{scaling_cumulant_kac}) given in the text. We start from the general formula valid
for all $n$ \cite{bendat} (see also \cite{bleher})
\bea
\langle N^2_n([0,x])\rangle = \int_0^x dt \rho_n(t) + \int_0^x dt_1  \int_0^x dt_2 {\cal K}_{n,2}(t_1,t_2) \;,\label{gen_formula_cumul_k}
\eea
where ${\cal K}_{n,2} (t_1,t_2)$ is the $2$-point correlation function of real roots of $K_n(x)$, given by
\bea
{\cal K}_{n,2} (t_1,t_2) = \frac{1}{4\pi^2 \sqrt{\det \Delta_n(t_1, t_2)}} \int_{-\infty}^\infty \int_{-\infty}^\infty |y_1 y_2| 
e^{-\tfrac{1}{2}(Y \Omega_n(t_1,t_2),Y)  } dy_1 dy_2 \;, \label{def_Kn}
\eea
where $Y = (y_1, y_2)$, $\Delta_n(t_1,t_2)$ is the $4\times 4$ covariance matrix of $(K_n(t_1), K_n'(t_1),K_n(t_2),K'_n(t_2))$ and $\Omega_n(t_1,t_2)$ is the $2 \times 2$ matrix obtained by removing the first and
the third rows and columns from $\Delta_n(t_1,t_2)^{-1}$. In the limit $0<1-t_1\ll1$, $0<1-t_2\ll1$ and $n \to \infty$, keeping $u_1=n(1-t_1)$, $u_2=n(1-t_2)$ fixed one has
\bea
&&\langle K_n(t_1) K_n(t_2) \rangle = n g(u_1+u_2) \; , \; g(x) = \frac{1-e^{-x}}{x} \;,\nonumber \\
&&\langle K_n(t_1) K'_n(t_2) \rangle =  \langle K'_n(t_1) K_n(t_2) \rangle = - n^2 g'(u_1+u_2) \;, \nonumber \\
&&\langle K'_n(t_1) K'_n(t_2) \rangle =  n^3 g''(u_1+u_2) \;. \label{correlators} 
\eea
Using these relations (\ref{correlators}), one obtains the matrix $\Delta_n(t_1,t_2)$ in this scaling limit as 
\begin{eqnarray}\label{struct_delta}
\Delta_n(t_1,t_2) = 
\begin{pmatrix}
n\,g(2u_1)&-n^2\,g'(2u_1)&n\,g(u_1+u_2)&-n^2\,g'(u_1+u_2)\\
-n^2\,g'(2u_1)&n^3\,g''(2u_1)&-n^2\,g'(u_1+u_2)&n^3\,g''(u_1+u_2)\\
n\,g(u_1+u_2&-n^2\,g'(u_1+u_2)&n\,g(2u_2)&-n^2\,g'(2u_2)\\
-n^2\,g'(u_1+u_2)&n^3\,g''(u_1+u_2)&-n^2\,g'(2u_2)&n^3\,g''(2u_2)
\end{pmatrix} \;,
\end{eqnarray}
from which one gets
\bea
&&\det  \Delta_n(t_1,t_2) = n^8 \det \tilde \Delta(u_1,u_2) \;,\label{scaling_det}\\
&&\tilde \Delta(u_1,u_2) = 
\begin{pmatrix}
g(2u_1)&-g'(2u_1)&g(u_1+u_2)&-g'(u_1+u_2)\\
-g'(2u_1)&g''(2u_1)&-g'(u_1+u_2)&g''(u_1+u_2)\\
g(u_1+u_2&-g'(u_1+u_2)&g(2u_2)&-g'(2u_2)\\
-g'(u_1+u_2)&g''(u_1+u_2)&-g'(2u_2)&g''(2u_2)
\end{pmatrix} \;. \label{delta_tilde} 
\eea
From the structure of $\Delta_n(t_1,t_2)$ in the scaling limit (\ref{struct_delta}), one obtains the one of $\Omega_n(t_1,t_2)$ in Eq. (\ref{def_Kn}) as
\bea
\Omega_n(t_1,t_2) = n^{-3} \tilde \Omega(u_1,u_2) \;,\label{scaling_omega}\\
\tilde \Omega(u_1,u_2) = 
\begin{pmatrix}
A(u_1,u_2)&B(u_1,u_2)\\
B(u_1,u_2)&C(u_1,u_2)
\end{pmatrix} \;, \label{tilde_omega}
\eea
where $\tilde \Omega(u_1,u_2)$, is obtained by removing the first and the third rows and columns from $\tilde \Delta(u_1,u_2)^{-1}$. The functions $A(u_1,u_2), B(u_1,u_2), C(u_1,u_2)$ have complicated expressions not given
here which can be easily computed {\it e.g.} using Mathematica. Using the scaling forms (\ref{scaling_det}, \ref{scaling_omega}) one obtains ${\cal K}_{n,2}(t_1,t_2)$ in Eq. (\ref{def_Kn}) in the scaling limit as
\bea
&&{\cal K}_{n,2}(t_1,t_2) = n^2 \tilde {\cal K}(u_1,u_2) \label{scaling_K}\\
&&\tilde {\cal K} = \frac{(\det \tilde \Delta)^{-{1}/{2}}} { \pi^2AC(1-(\delta)^2)} \left(1+\frac{\delta}{\sqrt{1-(\delta)^2}} \ArcSin\delta \right)  \;, \; \delta=\frac{B}{\sqrt{A\,C}} \;. \label{expr_ktilde}
\eea
where we used the notation $\tilde \Delta \equiv \tilde \Delta(u_1,u_2)$ and the expression \cite{bleher} (see also Ref.~\cite{bendat})
\bea\label{abs_integration}
\int_{-\infty}^\infty \int_{-\infty}^{\infty} |y_1 y_2| e^{-\tfrac{1}{2}(Ay_1^2 + 2By_1y_2 + Cy_2^2)} dy_1 dy_2 = \frac{4}{AC(1-(\delta)^2)} \left(1+\frac{\delta}{\sqrt{1-(\delta)^2}} \ArcSin\delta \right)
\eea
Using this scaling form (\ref{scaling_K}), together with the one for the density in Eq. (\ref{scaling_density_kac}), one obtains $\langle N_n([0,x])^2\rangle_c$ in the limit $0<(1-x)\ll 1$, with $n \to \infty$ keeping $y=n(1-x)$ fixed as 
\begin{eqnarray}
\langle N_n([0,x])^2\rangle_c  \sim \int_y^n \rho^K(u) du + \int_y^n du_1\int_y^n du_2 \,(\tilde {\cal K}(u_1,u_2) -\rho^K(u_1) \rho^K(u_2)) \;.  \label{scaling_cumulant_eq1} 
\end{eqnarray}
Alternatively, one can write Eq. (\ref{scaling_cumulant_eq1}) in the large $n$ limit as
\begin{eqnarray}
&&\langle N_n([0,x])^2\rangle - \langle N_n([0,1])^2\rangle = -\nu_-(n(1-x)) \\
&&\nu_-(y) = \int_0^y du \rho^K(u) +2 \int_0^y du_1 \int_{u_1}^\infty du_2 \,( \tilde {\cal K}(u_1,u_2) - \rho^K(u_1) \rho^K(u_2)) \;. \label{expr_g}
\end{eqnarray}

\subsection{Large argument behavior of $\nu_-(y)$}

To compute $\nu_-(y)$ in the large $y$ limit, one compute $d{\nu_-}(y)/dy$ from Eq. (\ref{expr_g})
\bea
\frac{d \nu_-(y)}{dy} = \rho^K(y) + 2 \int_y^\infty du_2 ( \tilde {\cal K}(y,u_2) - \rho^K(y) \rho^K(u_2)) \;. \label{expr_gprime}
\eea
The large $y$ behavior of $\rho^K(y)$ has been computed previously (\ref{density_large_uplus}). To extract the large $y$ behavior of $d{\nu_-}(y)/dy$, one thus needs to compute the behavior of $\tilde {\cal K}(u_1,u_2)$ for $u_1,u_2 \gg 1$. We first analyse the behavior $\tilde \Delta(u_1,u_2)$ in Eq.  (\ref{delta_tilde}) for $u_1,u_2 \gg 1$. This yields
\bea \label{delta_tilde_large}
\tilde \Delta(u_1,u_2) \sim
\begin{pmatrix}
\frac{1}{2u_1}&\frac{1}{4u_1^2}&\frac{1}{u_1+u_2}&\frac{1}{(u_1+u_2)^2}\\
\frac{1}{4u_1^2}&\frac{1}{4u_1^3}&\frac{1}{(u_1+u_2)^2}&\frac{2}{(u_1+u_2)^3}\\
\frac{1}{u_1+u_2}&\frac{1}{(u_1+u_2)^2}&\frac{1}{2u_2}&\frac{1}{4u_2^2}\\
\frac{1}{(u_1+u_2)^2}&\frac{2}{(u_1+u_2)^3}&\frac{1}{4u_2^2}&\frac{1}{4u_2^3}
\end{pmatrix} \;,
\eea
from which one gets
\beq
\det \tilde \Delta(u_1,u_2) \sim \frac{1}{256 (u_1 u_2)^4} \left( \frac{u_1-u_2}{u_1+u_2}  \right)^8 \;.
\eeq
From Eq. (\ref{delta_tilde_large}), one gets the behavior of the matrix $\tilde \Omega$ in Eq. (\ref{tilde_omega}) as 
\bea
&&A(u_1,u_2) \sim 8 u_1^3 \left(\frac{u_1+u_2}{u_1-u_2} \right)^4  \label{a_large} \\
&&B(u_1,u_2) \sim 16 \frac{u_1^2 u_2^2}{u_1+u_2} \left(\frac{u_1+u_2}{u_1-u_2} \right)^4 \label{b_large} \\
&&C(u_1,u_2) \sim 8 u_2^3 \left(\frac{u_1+u_2}{u_1-u_2} \right)^4 \label{c_large}
\eea
which gives $\delta= B/\sqrt{AC}$ as
\beq
\delta(u_1,u_2) \sim \frac{2 \sqrt{u_1 u_2}}{u_1 +u_2} \label{delta_large}
\eeq
Finally, using Eq. (\ref{a_large}-\ref{delta_large}) together with Eq. (\ref{density_large_uplus}) in $d=2$ one gets
\bea
\tilde {\cal K}(u_1,u_2) - \rho^K(u_1) \rho^K(u_2) \sim -\frac{1}{\pi^2(u_1+u_2)^2} + \frac{1}{2\pi^2 \sqrt{u_1 u_2}} \frac{|u_1-u_2|}{(u_1+u_2)^2} \ArcSin{\frac{2\sqrt{u_1u_2}}{u_1+u_2}} \;.
\eea
Therefore, from Eq. (\ref{expr_gprime}), one gets
\beq
 \frac{d \nu_-(y)}{dy}\sim \frac{1}{y} \left( \frac{2}{\pi^2}-\frac{1}{\pi}\right) + {\cal O}(y^{-2}) \;, \label{gprime_large}
\eeq
where we have used the identity 
\beq
\int_1^\infty \frac{dv}{\sqrt{v}} \frac{v-1}{(v+1)^2} \ArcSin\left( \frac{2\sqrt{v}}{v+1}\right) = \frac{\pi}{2}-1 \;.
\eeq
Finally, one gets from Eq. (\ref{gprime_large}) that
\beq
\nu_-(y) \sim \left( \frac{2}{\pi^2}-\frac{1}{\pi}\right)  \log{y} + {\cal O}(1) \;.
\eeq

\section{Computation of $\langle N^2_n([-x,x])\rangle_c$ for Weyl
  polynomials}\label{appendix_scaling_weyl} 

\subsection{Scaling form}

In this appendix we give the detail of the computation of $\langle N_n([-x,x])^2 \rangle$ which leads to formula (\ref{scaling_cumulant_weyl}) in the text. Here again we start from the general formula valid for all $n$, similarly to Eq. (\ref{gen_formula_cumul_k})
\bea
\langle N^2_n([-x,x])\rangle = \int_{-x}^x dt \rho_n(t) + \int_{-x}^x dt_1  \int_{-x}^x dt_2 {\cal W}_{n,2}(t_1,t_2) \;, \label{gen_formula_cumul_w}
\eea
where ${\cal W}_{n,2} (t_1,t_2) $ is the $2$-point correlation function of the real roots of $W_n(x)$, given by formula (\ref{def_Kn}) where $K_n(x)$ is replaced by $W_n(x)$:
\bea
{\cal W}_{n,2} (t_1,t_2) = \frac{1}{4\pi^2 \sqrt{\det \Delta_n(t_1, t_2)}} \int_{-\infty}^\infty \int_{-\infty}^\infty |y_1 y_2| 
e^{-\tfrac{1}{2}(Y \Omega_n(t_1,t_2),Y)  } dy_1 dy_2 \;, \label{def_Wn}
\eea
where $Y = (y_1, y_2)$, $\Delta_n(t_1,t_2)$ is the $4\times 4$ covariance matrix of $(W_n(t_1), W_n'(t_1),W_n(t_2),W'_n(t_2))$ and $\Omega_n(t_1,t_2)$ is the $2 \times 2$ matrix obtained by removing the first and
the third rows and columns from $\Delta_n(t_1,t_2)^{-1}$. In the limit $n \to \infty$, keeping $t_1 <\sqrt{n}$, $t_2 < \sqrt{n}$ fixed one has
\bea
&&\langle W_n(t_1) W_n(t_2) \rangle = e^{t_1 t_2}  \;, \nonumber \\
&&\langle W'_n(t_1) W_n(t_2) \rangle = t_2 e^{t_1 t_2} \;, \nonumber \\
&&\langle W_n(t_1) W'_n(t_2) \rangle = t_1 e^{t_1 t_2} \;,\nonumber \\
&&\langle W'_n(t_1) W'_n(t_2) \rangle = (1+t_1 t_2) e^{t_1 t_2} \;, 
\eea
from which we obtain $\Delta_n(t_1,t_2) = \tilde \Delta(t_1,t_2)$ as
\bea
\tilde \Delta(t_1,t_2) = 
\begin{pmatrix}
e^{t_1^2} & t_1 e^{t_1^2} & e^{t_1 t_2} & t_1 e^{t_1 t_2} \\
t_1 e^{t_1^2}&(1+t_1^2) e^{t_1^2} & t_2 e^{t_1 t_2} & (1+t_1 t_2) e^{t_1 t_2}) \\
e^{t_1 t_2} & t_2 e^{t_1 t_2} & e^{t_2^2} & t_2 e^{t_2^2} \\
t_1 e^{t_1 t_2} & (1+t_1 t_2) e^{t_1 t_2} & t_2 e^{t_2^2} & (1+t_2^2)e^{t_2^2}
\end{pmatrix} \;. \label{delta_weyl}
\eea
The determinant is easily obtained as
\bea
\det \tilde \Delta(t_1,t_2) =  e^{(t_1+t_2)^2}\left( 4 \sinh^2{\left(\frac{(t_1-t_2)^2}{2} \right)} -(t_1-t_2)^4 \right)
\eea
From $\tilde \Delta(t_1,t_2)$ in Eq. (\ref{delta_weyl}), one obtains $\Omega_n(t_t,t_2) = \tilde \Omega(t_1,t_2)$ for large $n$ as
\bea
\tilde \Omega(t_1,t_2) = 
\begin{pmatrix}
D(t_1,t_2 & E(t_1,t_2)) \\
E(t_2,t_2) & F(t_1,t_2)
\end{pmatrix}
\eea
with
\bea
&&D(t_1,t_2) = \frac{e^{t_1^2+2 t_2^2}}{\tilde \Delta(t_1,t_2)} (1-e^{-(t_1-t_2)^2}(1+(t_1-t_2)^2) ) \;, \nonumber \\
&&E(t_1,t_2) = \frac{e^{3 t_1 t_2}}{\tilde \Delta(t_1,t_2)} (1-e^{(t_1-t_2)^2}(1-(t_1-t_2)^2)) \;, \nonumber\\
&&F(t_1,t_2) = \frac{e^{2 t_1^2+t_2^2}}{\tilde \Delta(t_1,t_2)} (1-e^{-(t_1-t_2)^2}(1+(t_1-t_2)^2) ) \;. \label{expr_abc}
\eea
Finally, using these expressions (\ref{expr_abc}) together with 
the formula in Eq. (\ref{abs_integration}), one obtains from Eq. (\ref{def_Wn}) that ${\cal W}_{n,2}(t_1,t_2) = \tilde{ \cal W}(t_1 - t_2)$ with
\bea
\tilde{ \cal W}(s)= \frac{1}{\pi^2} \frac{((1-e^{-s^2})^2-s^4e^{-s^2})^{\tfrac{1}{2}}}{1-e^{-s^2}} \left( 1 + \frac{\delta(s)}{\sqrt{1-(\delta(s))^2}} \ArcSin \delta(s)  \right) \;, \label{tilde_w}
\eea
with 
\beq
\delta(s) = \frac{e^{-s^2/2}(e^{-s^2}+s^2-1)}{1-e^{-s^2}-s^2 e^{-s^2}} \;.
\eeq
Notice that $\tilde {\cal W}(s)$ is the two point correlation function of real zeros of $W_n(x)$. Interestingly, this correlation function in Eq. (\ref{tilde_w}) coincides (up to a multiplicative prefactor $\pi^{-2}$) with the correlation of straightened zeros of $B_n(x)$ obtained in \cite{bleher}. Its asymptotic behaviors are given by
\begin{eqnarray}\label{asympt_tilde}
\tilde{ \cal W}(s) \sim 
\begin{cases}
\tfrac{|s|}{4 \pi} \;,\; s \to 0 \\
\tfrac{1}{\pi^2} + \tfrac{s^4 e^{-s^2}}{2\pi^2} \;,\; s \to \infty
\end{cases}
\end{eqnarray}

Finally, using the expression (\ref{gen_formula_cumul_w}) together with Eq. (\ref{def_Wn}) one has for $0< x <\sqrt{n}$, in the limit $n \to \infty$
\bea
&&\langle N^2_n([-x,x])\rangle_c = \nu(x) \;, \nonumber \\
&&\nu (x) = \frac{2 x}{\pi} + 2 \int_0^{2 x} ds (2x-s) (\tilde {\cal W}(s)-\pi^{-2}) \;. \label{g_tilde_weyl}
\eea

\subsection{Large argument behavior of $\nu(x)$}

To analyse the large $x$ behavior of $\nu(x)$, one computes $d\nu(x)/dx$ and gets immediately
\beq
\lim_{x \to \infty} \frac{d\nu(x)}{dx} = \frac{2}{\pi} + 4 \int_0^\infty ds (\tilde {\cal W}(s)-\pi^{-2})
\eeq
such that
\bea\label{large_g_weyl}
\nu(x) \propto 2 \nu_{\infty} x \;, \; x \gg 1
\eea
with $\nu_\infty = 0.181988...$, which leads to the value of $\theta_\infty$ up to order ${\cal O}(\epsilon^2)$
given in Eq. (\ref{proba_second_order_weyl}).

\section{Computation of $\langle N^3_n([a,b])\rangle_c$ for binomial
  polynomials}\label{appendix_scaling_bp} 

In this appendix, we give the detailed calculation of $\langle N^3_n([a,b])\rangle_c$ which leads to the formula (\ref{scaling_cumulant_bp_2}) given in the text, for $m=3$. We start with the general formula (see for instance Ref.~\cite{bleher}):
\bea
&&\langle N^3_n([a,b]) \rangle_c = \langle N_n[a,b]\rangle + 3 \langle N^2_n[a,b] \rangle_c \label{cumul_3_1}\\
&&+ \int_a^b dt_1 \int_a^b dt_2 \int_a^b dt_3 \left({\cal B}_{n,3}(t_1,t_2,t_3) - 3 {\cal B}_{n,2}(t_1,t_2) \rho_n(t_3) +  \rho_n(t_1) \rho_n(t_2)\rho_n(t_3)\right) \;, \label{cumul_3_2}
\eea
where ${\cal B}_{n,m}$ is the $m$-point correlation function of real roots of $B_n(x)$. In Eq. (\ref{cumul_3_1}), ${\cal B}_{n,2}(t_1,t_2)$ is given by the formula (\ref{def_Kn}) where $K_n(x)$ is replaced by $B_n(x)$ and ${\cal B}_{n,3}(t_1,t_2,t_3)$ is formally given by (see for instance Ref.~\cite{bleher})
\beq
{\cal B}_{n,3}(t_1,t_2,t_3) = \int_{-\infty}^\infty dy_1 \int_{-\infty}^\infty dy_2 \int_{-\infty}^\infty dy_3 |y_1 y_2 y_3| D_{n,3}(0,y_1,0,y_2,0,y_3;t_1,t_2,t_3)  \;,
\eeq
where $D_{n,3}(x_1,y_1,x_2,y_2,x_3,y_3)$ is the joint distribution density of $(B_n(t_1),B'_n(t_1),   (B_n(t_2),B'_n(t_2),(B_n(t_3),B'_n(t_3))$. According to Eq. (\ref{scaling_cumulant_bp_1}), the two terms in (\ref{cumul_3_1}) have the desired form (\ref{scaling_cumulant_bp_2}) for large $n$. To study the triple integral in Eq. (\ref{cumul_3_2}) in the large $n$ limit, we will use the results obtained in Ref.~\cite{bleher} :
\bea\label{result_bleher_2}
\lim_{n \to \infty} \left [\frac{{\cal B}_{n,2}(t_1,t_2)}{\rho_n(t_1) \rho_n(t_2)} \right]_{t_i =\langle N[0,s_i] \rangle^{-1}}=b_2(s_1,s_2) \equiv b_2(s_1-s_2) \;, 
\eea
where $\langle N[0,s_i] \rangle^{-1}$ is the inverse function of $\langle N[0,s_i] \rangle$. Similarly
\bea\label{result_bleher_3}
\lim_{n \to \infty} \left [\frac{{\cal B}_{n,3}(t_1,t_2,t_3)}{\rho_n(t_1) \rho_n(t_2)\rho_n(t_3)} \right]_{t_i =\langle N[0,s_i] \rangle^{-1}}=b_3(s_1,s_2,s_3) \equiv b_3(s_1-s_2,s_2-s_3) \;,
\eea
where $b_{2}(u)$ and $b_3(u,v)$ are well defined functions, with a quite complicated expression not given here (see Ref.~\cite{bleher} for more detail). Given these results (\ref{result_bleher_2}, \ref{result_bleher_3}), it is natural to 
perform the change of variable $s_i = \langle N_n[0,t_i] \rangle$ in Eq. (\ref{cumul_3_2}), which yields
\bea
&& \int_0^x dt_1 \int_0^x dt_2 \int_0^x dt_3 \left({\cal B}_{n,3}(t_1,t_2,t_3) - 3 {\cal B}_{n,2}(t_1,t_2) \rho_n(t_3) +  \rho_n(t_1) \rho_n(t_2)\rho_n(t_3)\right) \\
&& = \int_{\langle N_n[0,a] \rangle}^{\langle N_n[0,b] \rangle} ds_1 \int_{\langle N_n[0,a] \rangle}^{\langle N_n[0,b] \rangle} ds_2 \int_{\langle N_n[0,a] \rangle}^{\langle N_n[0,b] \rangle} ds_3 \left( b_3(s_1-s_2,s_2-s_3)-3 b_2(s_1-s_2) +2 \right) \;.
\eea
Performing the change of variables $s_i \to s_i - \langle N_n[0,a] \rangle$, one has
\bea\label{cumu_3_2_scaled}
&& \int_0^x dt_1 \int_0^x dt_2 \int_0^x dt_3 \left({\cal B}_{n,3}(t_1,t_2,t_3) - 3 {\cal B}_{n,2}(t_1,t_2) \rho_n(t_3) +  \rho_n(t_1) \rho_n(t_2)\rho_n(t_3)\right) \\
&& = \int_{0}^{\langle N_n[a,b] \rangle} ds_1 \int_{0}^{\langle N_n[a,b] \rangle} ds_2 \int_{0}^{\langle N_n[a,b] \rangle} ds_3 \left( b_3(s_1-s_2,s_2-s_3)-3 b_2(s_1-s_2) +2 \right) \;.
\eea
Given that $\langle N_n[a,b] \rangle \propto \sqrt{n}$ in the large $n$ limit for $a,b > n^{-1/2}$, 
 one gets the multiple integral in Eq. (\ref{cumu_3_2_scaled}) in the large $n$ limit as  
\bea
&&\int_0^{\langle N_n[a,b] \rangle} ds_1 \int_0^{\langle N_n[a,b] \rangle} ds_2 \int_0^{\langle N_n[a,b] \rangle} ds_3 \left( b_3(s_1-s_2,s_2-s_3)-3 b_2(s_1-s_2) +2 \right) \nonumber \\
&& \sim \sigma \langle N_n[a,b] \rangle \;, \label{cumul_3_final}
\eea
with $\sigma = 3 \int_{-\infty}^0 du \int_{-\infty}^0 dv \left(b_3(u,v)-3b_2(v)+2 \right)$, 
provided this double integral over $u,v$ is well defined (which we can only assume here given the complicated expression of $b_3(u,v)$).

Finally, combining Eq. (\ref{cumul_3_1}, \ref{cumul_3_2}, \ref{scaling_cumulant_bp_1}) together with Eq. (\ref{cumul_3_final}) one obtains that
\beq
\langle N^3_n[a,b]\rangle_c \propto \beta_3 \langle N_n[a,b]\rangle \;,
\eeq
with $a_3 = 1 + 3a_2 +\sigma$. Notice that the crucial point in the derivation of this relation is the fact that  $\langle N_n[a,b] \rangle \propto \sqrt{n}$ for
any fixed $a,b > n^{-1/2}$. One expects a similar mechanism to hold for higher values of $m$, yielding Eq.  (\ref{scaling_cumulant_bp_2}) in the text.
\beq
\langle N^m_n[a,b]\rangle_c \propto \beta_m \langle N_n[a,b]\rangle \;.
\eeq

\end{appendix}

\end{document}